%% file: main.tex
\documentclass[letterpaper,twocolumn,10pt]{article}
\usepackage{usenix,epsfig,endnotes}
\usepackage{cite}
\usepackage[numbers]{natbib}
\usepackage{amsmath,amssymb,amsfonts}
\usepackage{graphicx}
\usepackage{textcomp}
\usepackage{xcolor}  
\usepackage[utf8]{inputenc}
\usepackage{newunicodechar}
\newunicodechar{▸}{\texttt{\textgreater}}
\usepackage{amsthm} 
\usepackage{natbib}
\usepackage{soul}
\usepackage{multirow}
\usepackage{xcolor}
\usepackage{hyperref}
\usepackage{cite}
\usepackage{amsmath,amssymb,amsfonts}
\usepackage{booktabs}
\usepackage{multirow}
\usepackage{cleveref}
\usepackage{caption}
\usepackage{hyperref}
\usepackage{tcolorbox} 
\usepackage{textgreek}
\tcbuselibrary{theorems} 
\tcbuselibrary{theorems, skins, breakable}
\tcbuselibrary{listings} 
\usepackage{amsmath}
\usepackage{algorithm}
\usepackage{algpseudocode}
\usepackage{caption}
\usepackage{colortbl}
\usepackage{xspace}
\usepackage{tcolorbox}

\newtcolorbox[auto counter]{observation}[2][]{colback=gray!5!white,colframe=gray!20!black,fonttitle=\bfseries, title=\textbf{Observation~\thetcbcounter~(triangle)}: #1, #2, coltitle=black, sharp corners, boxrule=0.5mm}

\newtcolorbox[auto counter]{takeaway}[2][]{colback=red!10, colframe=red!60, fonttitle=\bfseries, title=Takeaway~\thetcbcounter: #1, #2}

\newtcolorbox[auto counter]{game}[2][]{
    enhanced,
    colback=white,
    colframe=white,                 
    coltitle=black,
    colbacktitle=white,
    fonttitle=\bfseries,
    sharp corners,
    boxrule=0pt,                    
    titlerule=0pt,                  
    left=5pt, right=5pt,
    top=5pt, bottom=5pt,
    before skip=15pt,
    after skip=15pt,
    title=Game~\thetcbcounter: #2,
    label={#1},
    overlay unbroken={
        \draw[line width=1pt, black]
            (frame.south west) rectangle (frame.north east);
        \draw[line width=1pt, black]
            ([xshift=0pt]frame.north west |- title.south) --
            ([xshift=0pt]frame.north east |- title.south);
    }
}

\definecolor{commentcolor}{rgb}{0.5, 0.5, 0.5}
\definecolor{myorange}{RGB}{255, 165, 0}  
\definecolor{mygreen}{RGB}{60, 179, 113}   
\definecolor{myblue}{RGB}{70, 130, 180}    
\definecolor{mypink}{RGB}{246, 94, 144}    
\definecolor{mygray}{RGB}{73, 72, 73}

\makeatletter

\def\BibTeX{{\rm B\kern-.05em{\sc i\kern-.025em b}\kern-.08em
    T\kern-.1667em\lower.7ex\hbox{E}\kern-.125emX}}

\usepackage{cite}
\usepackage[numbers]{natbib}
\usepackage{amsmath,amssymb,amsfonts}
\usepackage{graphicx}
\usepackage{textcomp}
\usepackage{xcolor}
\usepackage{amsthm}
\usepackage{soul}
\usepackage{multirow}
\usepackage{hyperref}
\usepackage{booktabs}
\usepackage{cleveref}
\usepackage{caption}
\usepackage{subcaption}
\usepackage{microtype}
\usepackage{pifont}
\usepackage{wasysym}
\usepackage{enumitem}
\usepackage{tikz}
\usepackage{tcolorbox}
\usepackage{textgreek}
\usepackage{algorithm}
\usepackage{algpseudocode}
\usepackage{pifont} 
\usepackage{fontawesome}

\hypersetup{
    urlcolor= [rgb]{0.2314,0.1922,0.5804}  
}

\definecolor{commentcolor}{rgb}{0.5, 0.5, 0.5}
\definecolor{myorange}{RGB}{255, 165, 0}
\definecolor{mygreen}{RGB}{60, 179, 113}
\definecolor{myblue}{RGB}{70, 130, 180}
\definecolor{mypink}{RGB}{246, 94, 144}
\definecolor{mygray}{RGB}{73, 72, 73}

\newlist{circlelist}{enumerate}{1}
\setlist[circlelist,1]{label=\protect\circled{\arabic*}, leftmargin=*}
\newcommand*\circled[1]{\tikz[baseline=(char.base)]{
            \node[shape=circle,draw,inner sep=0.7pt,fill=black, text=white, font=\footnotesize] (char) {#1};}}

\tcbset{
  questionbox/.style={
    colback=white,
    colframe=black,
    boxrule=0.4pt,
    arc=0pt,
    left=6pt,
    right=6pt,
    top=6pt,
    bottom=6pt,
    sharp corners,
    before skip=10pt,
    after skip=10pt,
    breakable
  }
}

\newcommand{\sys}{\mbox{\textsc{RULI}}\xspace}
\title{\LARGE \bf Rectifying Privacy and Efficacy Measurements in Machine Unlearning: A New Inference Attack Perspective}
\usepackage[T1]{fontenc}
\usepackage{usenix}

\begin{document}

\author{
{\rm Nima Naderloui$^{1}$, Shenao Yan$^1$, Binghui Wang$^2$, Jie Fu$^3$, Wendy Hui Wang$^3$, Weiran Liu$^4$, Yuan Hong$^1$}\\
\textit{$^1$University of Connecticut, $^2$Illinois Institute of Technology}, \\\textit{$^3$Stevens Institute of Technology, $^4$Alibaba Group}
}

\maketitle

\input{Abstract}
\input{Introduction}

\input{Background}

\input{Framework}

\input{Evaluation}

\input{Discussion}

\input{conclusion}

{\small
\bibliographystyle{plainurl}  
\bibliography{references}
}

\appendix
\input{Appendix}

\end{document}

%% file: Abstract.tex
\begin{abstract}
Machine unlearning focuses on efficiently removing specific data from trained models, addressing privacy and compliance concerns with reasonable costs. Although exact unlearning ensures complete data removal equivalent to retraining, it is impractical for large-scale models, leading to growing interest in inexact unlearning methods. However, the lack of formal guarantees in these methods necessitates the need for robust evaluation frameworks to assess their privacy and effectiveness. In this work, we first identify several key pitfalls of the existing unlearning evaluation frameworks, e.g., focusing on average-case evaluation or targeting random samples for evaluation, incomplete comparisons with the retraining baseline. Then, we propose \sys (\emph{\underline{R}ectified \underline{U}nlearning Evaluation Framework via \underline{L}ikelihood \underline{I}nference}), a novel framework to address critical gaps in the evaluation of inexact unlearning methods. \sys introduces a dual-objective attack to measure both unlearning efficacy and privacy risks at a per-sample granularity. Our findings reveal significant vulnerabilities in state-of-the-art unlearning methods, where \sys achieves higher attack success rates, exposing privacy risks underestimated by existing methods. Built on a game-based foundation and validated through empirical evaluations on both image and text data (\emph{spanning tasks from classification to generation}), \sys provides a rigorous, scalable, and fine-grained methodology for evaluating unlearning techniques.\footnote{Code is available at~\url{https://github.com/datasec-lab/Ruli}}

\end{abstract}

%% file: Introduction.tex
\section{Introduction} 

Unlearning is crucial in modern machine learning, enabling the efficient removal of specific data (unlearn samples) or knowledge from trained models. It ensures compliance with privacy laws~\cite{GDPR2016, CCPA2018}, corrects outdated~\cite{ginart2019making} or harmful content~\cite{bender2021dangers}, and keeps models ethical and adaptable. As models and datasets continue to scale, interest in machine unlearning~\cite{bourtoule2021machine, adaptive_unlearning, thudi2022unrolling} has grown rapidly. While retraining from scratch without the removed samples provides a theoretically sound solution, it remains computationally expensive. To address this, a variety of inexact unlearning methods~\cite{bourtoule2021machine, guo2019certified} have been proposed as more efficient alternatives. Recent efforts focus on improving the efficiency, robustness, and accuracy of these approaches. Accordingly, rigorous evaluation is critical to ensure their practical effectiveness and to support trustworthy deployment by model providers and users alike. 

First, from a privacy perspective, techniques such as differentially private training~\cite{dpsgd} and defenses against inference attacks~\cite{nasr2018machine, SELENA_SELF_MIA,FengMHYKW025} typically require strong empirical validation—often through \emph{inference attacks} (e.g., membership inference attacks~\cite{nasr2023tight, aerni2024evaluations, panda2025privacy})—to assess their effectiveness in mitigating privacy risks under worst-case or average-case scenarios. Substantial research efforts have been dedicated to the design and evaluation of privacy mechanisms, often supported by empirical studies using inference attacks. 
Inexact unlearning can also be considered as a post-hoc privacy mechanism to efficiently remove samples and therefore, selectively~\cite{golatkar2020eternal} protect the privacy for a portion of data upon requests. Similarly, we require strong verifications to evaluate and minimize the privacy leakage for any unlearned sample~\cite{jagielski2022measuring, hayes2024inexact}. Existing unlearning works often report high protection against privacy attacks~\cite{sparse-mia, certified2, scrub}. However, Hayes et al. initiated efforts to advance stronger sample-level attacks~\cite{hayes2024inexact} indicating stronger attackers are required to evaluate unlearning. According to~\cite{hayes2024inexact}, existing algorithms are more vulnerable to their per-sample~\cite{carlini_lira} adapted attack. These insights motivate us to further investigate the following questions.

\vspace{0.05in}
\textit{Q1 [\textbf{Privacy}]: Can existing inference attacks accurately measure the privacy leakage from unlearned models?} 
\vspace{0.05in}

Second, another key requirement for evaluating unlearning success is assessing how closely it approximates the gold standard of retraining~\cite{acl_efficacy, hayes2024inexact}.

\vspace{0.05in}
\textit{Q2 [\textbf{Efficacy}]: How to accurately measure the difference between the unlearned model and the retrained model?} 
\vspace{0.05in}

Since efficacy reflects whether the requested samples have been effectively removed from the model through unlearning, it is related to (but distinct from) the privacy of the unlearned samples. While privacy concerns the risk of information leakage (``existence of unlearned samples''), efficacy focuses on the actual removal of the samples from the model.

\subsection{Contributions}

Motivated by the above questions, this work contributes the following to the field of machine unlearning.

\vspace{0.05in}

\noindent\textbf{(1) Identifying the limitations of existing inference attacks in evaluating unlearning}. We observe that current inference attacks on unlearning methods~\cite{sparsity, hayes2024inexact, fan2023salun} might not sufficiently challenge unlearning algorithms to enable a comprehensive evaluation of their effectiveness. First, most evaluations typically focus on \emph{average-case} scenarios and targeting \emph{random} samples, neglecting the \emph{per-sample} vulnerabilities associated with specific high-risk (or particularly vulnerable) data samples~\cite{onion}. Second, we also observe that simply comparing the retrained model with the unlearned model's accuracy may not yield an accurate assessment of privacy leakage~\cite{hayes2024inexact}.

\vspace{0.05in}

\noindent\textbf{(2) Developing novel inference attacks for dual measurements on privacy leakage and efficacy}. To address these deficiencies, we revisit the theoretical foundations of adversarial settings and design novel membership inference attacks (MIAs) \cite{shokri2017membership, carlini_lira, choquette2021label} with two key objectives:

\vspace{-0.05in}

\begin{itemize}
    \item \textit{Privacy Leakage}: an MIA to assess whether a sample has been unlearned or was never part of the training set (and subsequently not unlearned) based solely on inferring the unlearned model. 

\vspace{-0.05in}

    \item \textit{Efficacy}: an MIA to assess whether a sample's inference output corresponds to an unlearned or retrained model. 
\end{itemize}

\vspace{-0.05in}

With new MIAs, we propose a novel unlearning evaluation framework \sys (\emph{\underline{R}ectified \underline{U}nlearning Evaluation Framework via \underline{L}ikelihood \underline{I}nference}). To our best knowledge, \sys takes the first step to perform per-sample, targeted attacks on vulnerable samples in unlearning using refined membership signals and hypothesis testing, enabling the evaluation of both efficacy and privacy leakage at a reasonable attack cost.

\vspace{0.05in}

\noindent\textbf{(3) High attack performance and accurate unlearning evaluation.} We evaluate \sys against state-of-the-art (SOTA) attacks on standard unlearning benchmarks and conduct a comprehensive study of how different target samples exhibit varying levels of privacy leakage. For example, using a Gradient Ascent-based unlearning method, \sys achieves at least 20\% TPR@1\% FPR on CIFAR-10 and CIFAR-100, and up to 54\% TPR@1\% FPR when unlearning targeted 7-gram sequences from the WikiText-103 dataset. Our results also demonstrate that existing average-case attacks~\cite{sparsity, hayes2024inexact, fan2023salun} and advanced methods like U-LiRA~\cite{hayes2024inexact} have substantially underestimated the privacy leakage (and efficacy) especially under target random samples. 

\begin{figure}[!h]
    \centering \includegraphics[width=0.9\linewidth]{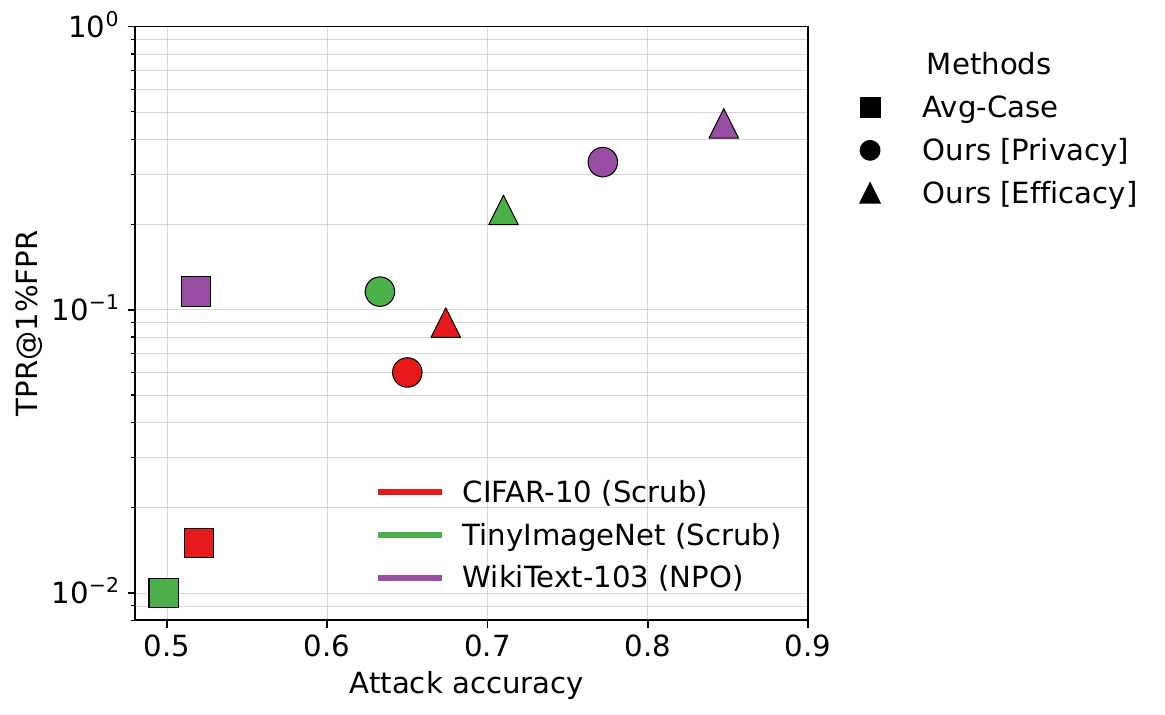}\vspace{-0.1in}
    \caption{\sys (ours) separates privacy leakage and unlearning efficacy through targeted membership inference. Population average-case attacks~\cite{graves2021amnesiac} consistently \emph{underestimate privacy leakage} and \emph{fail to capture unlearning efficacy}. This figure demonstrates some representative results, e.g., Scrub~\cite{scrub} on CIFAR-10 (ResNet-18) and TinyImageNet (Vision Transformer~\cite{liu2021swin}) for image sample unlearning, and NPO~\cite{zhang2024negative-NPO} on WikiText-103 (GPT2-small~\cite{radford2019language}) for token sequence unlearning. U-LiRA~\cite{hayes2024inexact} falls between \sys and average-case attacks, as it targets random samples on a per-sample basis but considers the efficacy similar to the privacy leakage.}

    \vspace{-0.1in}
    \label{fig:mia_introduction}
\end{figure}

As ~\autoref{fig:mia_introduction} shows, we uncover significantly stronger privacy leakage and highlight that efficacy is a distinct metric from privacy. We find that most inexact unlearning methods cannot closely approximate retraining. In other words, inexact unlearning may fail to remove samples as effectively as retraining, and even strong unlearning methods may fail to protect vulnerable samples when injected as canaries. 

\vspace{-0.2in}

\noindent

\noindent\textbf{(4) Generalizability.} \sys can be generalized to broadly support diverse tasks (from image classification to text generation) in multiple domains. We have validated its effectiveness on both image datasets (e.g., CIFAR-10/100, TinyImageNet) and text datasets (e.g., WikiText-103) by unlearning targeted data samples from various models, such as images from ViT models and token sequences from language models.

%% file: Background.tex
\newtheorem{definition}{Definition}[section]
\section{Preliminaries}

\subsection{Machine Unlearning}
Machine Unlearning refers to the process of selectively forgetting specific pieces of data that a model has previously trained on.
Consider a model, denoted as $\mathcal{\theta_I}$, trained on a dataset $D_\text{train}$; any samples belonging to $D_\text{train}$ can be subjected to unlearning, often due to the \textit{right to be forgotten} (Article 17 of the GDPR)~\cite{GDPR2016, CCPA2018} and form a forget set $D_f$.
The challenge is to erase the influence of ${D}_f$ while preserving the utility of the remain data ${D}_r = {D}_\text{train}\backslash {D}_f$. 
Machine unlearning employs various techniques to efficiently remove the influence of ${D}_f$ from a model originally trained on ${D}_\text{train}$. The objective is for the resulting unlearned model to closely approximate a model trained exclusively on ${D}_r$, without requiring full retraining.

\vspace{0.05in}
\noindent\textbf{Unlearning Measurement}. Most inexact unlearning methods~\cite{sparsity, fan2023salun} consider accuracy measurements to find how much the accuracy on $D_r$, $D_f$, and also unseen test data ($D_\text{test}$) differs with retraining.  
Previous works have also explored the use of Membership Inference Attacks (MIAs)~\cite{sparsity, chen2021machine, chen2023boundary, yeom_mia_loss, song2021systematic} to evaluate unlearning, defining success based on the attacker's ability to distinguish whether a sample was unlearned or included in retraining. Unlearning measurement is specifically critical in inexact unlearning algorithms as these methods, despite being efficient and scalable, often do not provide formal guarantees on data removal~\cite{guo2019certified}. Therefore, requires robust evaluations (often by strong attacks~\cite{hayes2024inexact}) to empirically ensure the unlearning algorithm is comparable to a retrained model and ensures privacy.

\vspace{-0.15 in}
\subsection{Inexact Unlearning Criteria}

Apart from the first goal of unlearning to speed up the data removal process compared to retraining (efficiency), given an initially trained model \( \theta_\mathcal{I} \) and an unlearning algorithm \( \mathcal{U} \), the goal of inexact unlearning is to remove a subset \( D_f \subset D_\text{train} \) from the training data by meeting the following key criteria:

\vspace{0.05in}
\noindent\textbf{(1) Accuracy.}  The unlearned model ($\theta_{\mathcal{U}}$) should ideally behave similarly to a retrained model ($\theta_{\mathcal{R}}$), recalling that the retrained model is the model trained excluding the $D_f$.  As mentioned, the basic accuracy test is that $\text{Acc}(\theta_{\mathcal{U}})\approx \text{Acc}(\theta_{\mathcal{R}})$ on all samples from $D_f$, $D_r$, and test data.

\vspace{0.05in}

\noindent\textbf{(2) Efficacy.} A widely accepted definition of the goal of unlearning is to ensure that the distribution of an unlearned model becomes indistinguishable from the distribution of a retrained model, which is trained exclusively on the remained dataset ($D_r$) ~\cite{sekhari2021remember, ginart2019making}. As highlighted in~\cite{hayes2024inexact}, there is a need to develop a tool to empirically estimate this indistinguishability. 

\vspace{0.05in}

\noindent\textbf{(3) Privacy.} It focuses on ensuring that the unlearned model ($\theta_{\mathcal{U}}$) provides no residual information about the forget data, $D_f$. In this context, the objective is that an MIA on $D_f$ using $\theta_{\mathcal{U}}$ should fail to infer whether any sample from $D_f$ was part of the original training process and unlearned ~\cite{jagielski2022measuring}. This test emphasizes the privacy protection provided by the unlearning process, ensuring that no attacker can exploit the unlearned model to extract information about any forget sample.

\vspace{0.05in}

\noindent\textbf{Unlearning vs. MIA-resilience.} Unlike MIA-mitigation methods~\cite{nasr2018machine, relaxloss, SELENA_SELF_MIA, inf2guard}, which aim to protect the entire training set, unlearning provides \textit{selective} privacy on demand. It targets only the requested forget set $D_f$, leaving $D_r$ unaffected unless specified. While applying MIA-resilient training to $D_f$ may reduce privacy leakage, it fails to meet the efficacy requirements. This calls for efficacy measurements besides the privacy leakage to evaluate unlearning.


\section{Unlearning Framework and Threat Model} 

Similar to previous works (e.g., \cite{scrub,chen2023boundary,thudi2022unrolling,guo2019certified,sparsity}), the framework is divided into two phases: \textit{training} and \textit{unlearning}. 
In the \textit{training} phase, the model is trained without anticipating future \textit{unlearning requests}, without training assumptions~\cite{chourasia2023forget, adaptive_unlearning} or incorporating additional checkpoints~\cite{scrub, bourtoule2021machine}. When an \textit{unlearning request} is received, the unlearning phase is activated as a separate process, efficiently updating the model to produce ${\theta_\mathcal{U}}$ with minimal computational costs. 
Our approach adheres to this flexible framework, enabling seamless integration of nearly all ``inexact unlearning'' methods.

Specifically, our unlearning framework involves three main entities: (1) Users (whose data are used for training the model and who may request to have their data unlearned), (2) Model Trainer (who trains the model and executes the unlearning process), and (3) Model Recipient (who can interact with the model, potentially performing inference attacks to evaluate privacy risks~\cite{inversion_S&P2024, sok_mia, chen2021machine, petsunlearning, chourasia2023forget}).

\vspace{0.05in}

\noindent\textbf{Users.} 
Individuals or entities who contribute to the data used in the model training. In the unlearning setting, users may request the removal or modification of their data from the trained model, necessitating mechanisms to ensure that their data can be effectively unlearned. These requests can pertain to any type of data contributed during training. Henceforth, we will refer to the specific data contributor and entities who request unlearning as the \textit{users} for simplicity.

\vspace{0.05in}

\noindent\textbf{Model Trainer.} 
The entity is responsible for storing data, training the model, and executing unlearning. The model trainer ensures that data are securely stored and processes requested queries. In the context of unlearning, it must also handle data removal requests and update the model to exclude the specified data while maintaining model performance and respecting user privacy rights~\cite{GDPR2016,CCPA2018}. We consider an honest model trainer, who performs training and unlearning with no adversarial objective against users' privacy.

\vspace{0.05in}

\noindent\textbf{Attacker's Capabilities.}
We consider an attacker with black-box access to \emph{only} the final unlearned model. That is, the attacker can query the unlearned model but has no access to the original (pre-unlearning) model or to any intermediate model states during the unlearning process. This assumption differs from some prior works~\cite{inversion_S&P2024, chen2021machine}, which assume the attacker has access to both the original and unlearned models. In contrast, our threat model aligns with a more realistic deployment scenario: once unlearning is complete, only the final unlearned model is released. This assumption also reflects the nature of inexact unlearning methods, which directly modify the original trained model rather than retraining from scratch. Unlike retraining, which inherently produces intermediate models, inexact unlearning does not expose meaningful unlearning checkpoints throughout the process. Therefore, it is reasonable to assume that the unlearned model becomes available only after the unlearning process is fully completed. 

Moreover, following~\cite{carlini_lira, hayes2024inexact, shokri2017membership}, we assume the attacker has knowledge of the training and unlearning algorithms, and can therefore train and access \textit{shadow models} to support the inference attack.

\noindent\textbf{Unlearning Efficacy Evaluator Capabilities.} We treat the efficacy evaluator as an honest model trainer who empirically assesses how well unlearning approximates retraining. As shown in Section~\ref{MIA_subsection}, querying only the final unlearned model is insufficient for this purpose. Unlike an attacker, the evaluator audits unlearning fidelity, not privacy leakage. The evaluator interacts with models via black-box queries, avoiding parametric comparisons~\cite{golatkar2020eternal, thudi2022necessity}, since inexact unlearning does not follow the same gradient path as retraining, nor does it guarantee indistinguishability with the retrained model. Instead, we assess behavioral differences resulting from the removal of requested samples.

%% file: Framework.tex
\section{A New Inference Attack on Unlearning}
\label{section-IV} 
In this section, we examine privacy leakage and efficacy in unlearning by identifying pitfalls in existing methods and introducing a new inference attack (Sections~\ref{sec:game_unlearning} and \ref{p4u-mia}).

\begin{game}[game1]{Existing MIA for unlearning privacy}
\small
\footnotesize
1. The \textit{challenger} trains a model with $D_\text{train}\subseteq \mathcal{D}$ and gets $\theta_\mathcal{I}$. 

2. The \textit{challenger} unlearns $D_f\subset D_\text{train}$ to get the unlearned model $\theta_\mathcal{U}$. 

3. The \textit{challenger} flips a coin $c$:
\begin{itemize}
    \item If $c=\text{head}$, the challenger chooses a data point $z$ from $D_f$
    \item If $c=\text{tail}$, the challenger chooses a data point $z$ from $\mathcal{D}\backslash D_{\text{train}}$
\end{itemize}

4. The \textit{challenger} sends the selected data point $z$ to the adversary. 

5. Given the unlearned model $\theta_\mathcal{U}$, the \textit{adversary} queries $z$ to determine if it is in $D_{\text{train}}$ and guess $\hat{c}$ = \{head, tail\}; \textit{adversary} wins if $\hat{c}=c$.
\end{game}

\subsection{Pitfalls in Existing Unlearning Evaluation}
\label{sec:game_unlearning}

Unlearning privacy is measured by the attacker's inability to distinguish between unlearned and never-trained samples~\cite{hayes2024inexact, scrub, jagielski2022measuring, shi2024muse}. This critical problem can be formalized through a game-based framework \cite{sok_mia, yeom_mia_loss, mahloujifar2021membership, onion}, which simulates the interaction between a challenger, responsible for training and unlearning a model, and an adversary, who aims to infer membership. In the challenger-adversary framework for unlearning, the process begins with a \textbf{challenger} \({C}\), who trains a model \(\theta_{\mathcal{I}}\) using a training dataset sampled from data distribution $\mathcal{D}$ (\(D_{\text{train}} \subseteq \mathcal{D}\))  via a training algorithm \(\mathcal{A}\). After initial training, the challenger selectively unlearns a subset of data points, \(D_f \subset D_{\text{train}}\), by applying an unlearning algorithm \(\mathcal{U}\). The resulting model, after unlearning, is denoted as \(\theta_{\mathcal{U}}\). 
The adversary's (${A}$) attack settings are captured through different games, characterized by different types of access and tools. In this paper, these games can serve as the theoretical representation for evaluating the privacy leakage and efficacy of unlearning.\footnote{Similar to the instantiated MIAs, the adversary in the game is also capable of training and accessing the shadow models (this applies to all the games and inference attacks defined in this paper). }

\textbf{Game \ref{game1}} formalizes the existing MIAs for unlearning privacy. Specifically, it assesses the adversary's ability to distinguish between data points that have been unlearned and those that were never part of the training dataset. In this setup,  the adversary has neither prior knowledge of nor control over the data samples requested for unlearning. This game simulates a scenario where the adversary can only query \emph{random samples} provided by the challenger, serving as a baseline for measuring privacy leakage in unlearning under non-targeted attacks. However, existing evaluations (mostly built upon this game and the corresponding MIA) exhibit significant shortcomings.

\vspace{0.05in}
\label{pitfall}
\noindent \textbf{Pitfall I: Average-case MIAs Cannot Fully Disclose Unlearning Privacy.} Nearly all existing works~\cite{sparsity, fan2023salun, scrub,fan2025challenging} rely on MIAs based on \textbf{Game~\ref{game1}} to evaluate the privacy leakage of their unlearning methods. The most naive approach~\cite{sparsity, fan2023salun} involves training an MI-classifier on equal-sized training/testing datasets to measure the attacker's ability to identify an unlearned sample as part of the training. A refined approach~\cite{graves2021amnesiac} employs population attacks, where many shadow models are trained and unlearned, and an MI-classifier is trained on \textit{unlearn} and \textit{out} (excluded from training) populations for distinguishing these two cases. 
While this improves upon naive approaches, it still evaluates privacy leakage in terms of aggregate metrics, failing to consider the unique memorization status of individual samples. Since memorization varies per sample, unlearning should also be evaluated at the per-sample level for accurate privacy assessment.

\vspace{0.05in}
\noindent \textbf{Pitfall II: Evaluating \emph{Random} Samples Underestimates Unlearning Privacy.} Recent trends in machine learning privacy research emphasize identifying and evaluating vulnerable samples, rather than assessing the privacy of the entire dataset~\cite{aerni2024evaluations, whatui2024you}. However, to our best knowledge, nearly all unlearning methods only consider the case of \textit{random sample evaluation} for unlearning—choosing random samples uniformly across all classes or within a single class—without accounting for the specific memorization level of each sample.  
While recent works~\cite{scrub, hayes2024inexact} suggest that unlearning random samples within a single class poses the greatest challenge, even more extreme cases can further challenge unlearning. More importantly, we show that evaluating random samples tends to underestimate the privacy leakage in unlearning.

\vspace{0.05in}
\noindent \textbf{Pitfall III: Incomplete Comparisons with the \textit{Retrain} Baseline (Efficacy)}. Many inexact unlearning methods are compared to a retrained model~\cite{retrain} to ensure that the unlearned model behaves similarly to it. However, these comparisons are often limited to accuracy metrics~\cite{scrub, fan2023salun, chen2023boundary}, which might be inaccurate (see~\autoref{tab:scrub_NegGrad} and \autoref{fig:retrained_unlearned}). For example, a model showing close accuracy performance to the retrained model might unlearn data samples differently than the retrained model. We emphasize the need to evaluate how individual samples are unlearned relative to their retrained counterparts. While this pitfall does not directly relate to privacy, it highlights the limitations of existing methods in quantifying the similarity between unlearned and retrained models beyond simple accuracy comparisons.

\subsection{Avoiding Pitfalls Requires New Games}
\label{p4u-mia}
Previous studies have assessed the unlearning success by observing the attacker's inability to execute an inference attack~\cite{jagielski2022measuring, scrub} on the unlearned model. 
As discussed in \textbf{Pitfall I}, although many unlearning benchmarks have demonstrated \textit{average-case} MIA-resilience, our focus is on evaluating unlearning through sample-specific membership signals for identifying underestimated privacy leakage, in line with recent privacy evaluation literature~\cite{carlini_lira, zarifzadeh2024low, aerni2024evaluations, nasr2023tight}.


\noindent\textbf{Per-Sample Privacy Evaluation for Unlearning}. MIAs are typically evaluated using metrics such as the Area Under the Curve (AUC) and TPR@lowFPR over the entire training dataset. For example, in CIFAR-10, the target training data consists of 50,000 samples equally split between trained and not-trained samples~\cite{carlini_lira}. However, MIA performance and memorization behavior are not uniform across all samples~\cite{aerni2024evaluations, carlini_lira}.
As shown in ~\autoref{tab:vulnerable_num} (in the experiments), targeting smaller, randomly selected subsets of training data often leads to degraded attack performance~\cite{aerni2024evaluations}, likely due to the increased likelihood of these smaller subsets containing predominantly safe samples~\cite{onion}. 
Consequently, uniformly selected smaller subsets result in lower attack accuracy and reduced TPR@lowFPR. This challenge is particularly relevant in the context of unlearning. Similarly, the target set (i.e., $D_\text{target}$) of membership inference is typically a smaller fraction of the training set, often less than 10\%~\cite{bourtoule2021machine}. Additionally, most unlearning algorithms are not designed to handle the removal of large portions of data~\cite{fan2023salun}. Thus, evaluations should not only consider individual samples but also include challenging scenarios for unlearning algorithms, avoiding \textbf{Pitfall II}.

\vspace{-0.05in}
\begin{game}[game2]{Targeted MIA for unlearning privacy}
\footnotesize
1. The \textit{challenger} trains a model with $D_\text{train}\subseteq \mathcal{D}$ and gets $\theta_\mathcal{I}$. 

\textcolor{myblue}{2. The \textit{adversary} chooses a target set $D_{\text{target}}$ and sends to \textit{challenger}}.

3. The \emph{challenger} unlearns \textcolor{myblue}{$D_f \cup \{D_\text{train} \cap D_\text{target}\}$} to get the model $\theta_\mathcal{U}$.

4. The \emph{challenger} flips a coin $c$:

\begin{itemize}
    \item If $c=$ head, the challenger chooses a data point $z$ from $D_f \cap D_\text{target}$   
    \item If $c=\text{tail}$, the challenger chooses a data point $z$ from \textcolor{myblue}{\textbf{$D_{\text{target}}\backslash D_\text{train}$}}
\end{itemize}

5. The \textit{challenger} sends the selected data point $z$ to the adversary. 

6. Given the unlearned model $\theta_\mathcal{U}$, the \textit{adversary} queries $z$ to determine if it is in $D_{\text{train}}$ and guess $\hat{c}$ = \{head, tail\}; \textit{adversary} wins if $\hat{c}=c$.
\end{game}

\vspace{-0.1in}
\noindent\textbf{Targeted MIA for Evaluating Unlearning Privacy}. To address these challenges, our approach is introduced in \textbf{Game \ref{game2}} (new contents marked in blue), which extends the attack setting to allow targeted access to the unlearned model. In this setting, the adversary selects specific targeted samples $D_\text{target}$ for unlearning rather than relying on random samples. Despite this extended capability, the adversary operates in a \textit{targeted black-box} threat model~\cite{tramer2022truth, wen2024privacy}, lacking access to internal parameters, gradients, or embeddings and unable to manipulate the unlearning process directly. 

\vspace{-0.05in}
\begin{game}[game3]{MIA for unlearning efficacy}
\footnotesize

1. The \textit{challenger} trains a model with $D_\text{train}\subseteq \mathcal{D}$ and gets $\theta_\mathcal{I}$. 

\textcolor{myblue}{2. The \textit{adversary} chooses a target set $D_{\text{target}}$ and sends to \textit{challenger}}.

3. The challenger unlearns \textcolor{myblue}{$D_f \cup \{D_\text{train} \cap D_\text{target}\}$}  to get the model $\theta_\mathcal{U}$.

4. The challenger flips a coin $c$:
\textcolor{mypink}{
\begin{itemize}
    \item If $c=$ head, the challenger chooses a data point $z$ from $D_f \cap D_\text{target}$, and the query result will be given as $f_{\theta_\mathcal{U}}(\cdot)$
    \item If $c=\text{tail}$, the challenger chooses a data point $z$ from $D_{\text{target}}\backslash D_\text{train}$, and the query result will be given as $f_{\theta_\mathcal{I}}(\cdot)$
\end{itemize}}

5. The \textit{challenger} sends the selected data point $z$ to the adversary. 

6. \textcolor{mypink}{Given the query from queries $z$ as $f_\theta(\cdot)$, the \textit{adversary} determines if $z$ is in $D_f$ and guess $\hat{c}$ = \{head, tail\}; \textit{adversary} wins if $\hat{c}=c$.}
\end{game}

\noindent\begin{figure*}[ht]
    \centering    \includegraphics[width=\linewidth]{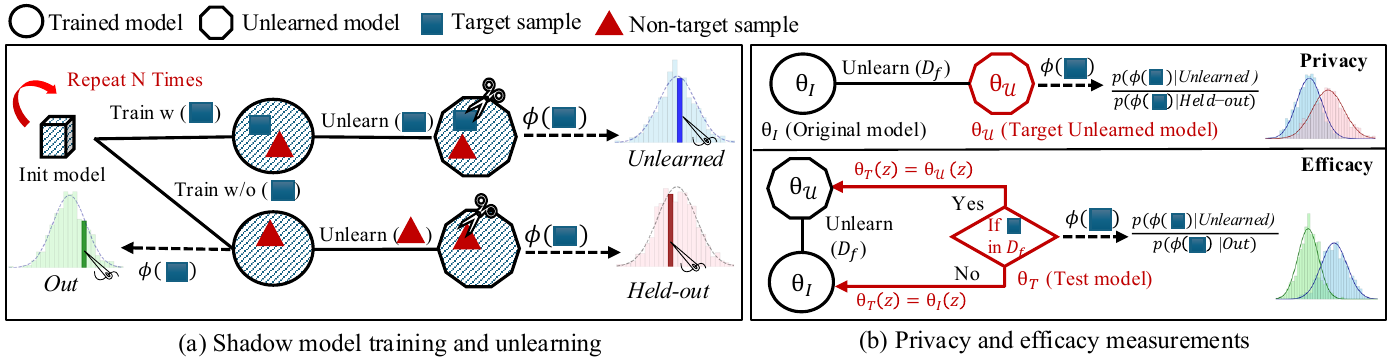}\vspace{-0.05in} 
    \caption{Overview of \sys shadow model training and unlearning to measure the privacy leakage and efficacy.}\vspace{-0.15in} 
    \label{fig:revision_framework}
\end{figure*}

\vspace{-0.1in}
\noindent\textbf{Evaluating Efficacy with Indistinguishability between Unlearned Model and Retrained Model}. 
As highlighted in \textbf{Pitfall III}, inexact unlearning lacks formal guarantees, making empirical evaluation essential~\cite{hayes2024inexact}. To assess whether an unlearning method approximates retraining behavior, we adopt a formal game formulation based on indistinguishability between the unlearned model and a retrained model. In \textbf{Game \ref{game3}} (new contents marked in pink), the challenger first trains and unlearns a model to obtain $\theta_{\mathcal{U}}$. The challenger flips a coin and chooses the target sample from the target set chosen and shared by the adversary. If the sample belongs to $D_f$, the query result will be given to the adversary as $f_{\theta}(\cdot) = f_{\theta_{\mathcal{U}}}(\cdot)$; and if the sample was not involved in training, the query result will be given as $f_{\theta}(\cdot) = f_{\theta_{I}}(\cdot)$. The adversary must guess if that sample is unlearned or retrained. 
\textit{This game is fair:} The adversary receives output from only one model and never has access to internal weights or both models simultaneously. The baseline success rate remains 50\% (random guessing), and the adversary gains no extra advantage. The game only tests distinguishability between 1) a sample that was unlearned, and 2) a sample that was never trained on, based solely on observable predictions.

We highlight that \textbf{Game~\ref{game2}} is a practical privacy attack for identifying the privacy leakage in the unlearned model $\theta_\mathcal{U}$. \textbf{Game~\ref{game3}} defines the foundation for an attack to differentiate the retrained model from an unlearned model (efficacy), and it is not instantiated as a real-world attack.

\subsection{\sys to Resolve Pitfalls}
\label{pifall_mitigation} 
\subsubsection{Theoretical Foundation for \sys}
Earlier, we have established the theoretical foundation to tackle Pitfalls. Now we design a new MIA for instantiating the attack from the empirical perspective. Similar to~\cite{carlini_lira}, we begin with the distributions an adversary requires to make a hypothesis test~\cite{carlini_lira} on them. According to \textbf{Game~\ref{game2}}, adversary requires two distinct distributions of unlearned models.

\vspace{-0.1in}

\begin{align*}
\mathcal{Q}_u =& \{\theta \leftarrow \mathcal{U}(\mathcal{A}(D_{\text{train}} \cup \{z\}), \{z\} \cup D_f \mid D_f\subset D_{\text{train}} \sim \mathcal{D})\}\nonumber\\
\mathcal{Q}_h = &\{\theta \leftarrow \mathcal{U}(\mathcal{A}(D_{\text{train}}), D_f \mid D_f\subset D_{\text{train}} \sim \mathcal{D})\}
\end{align*}

where $\mathcal{U}$ and $\mathcal{A}$ refer to the unlearning and training algorithms, respectively. Specifically, $\mathcal{Q}_u$ denotes the distribution of models in which the target sample ($z$) was included in training and subsequently unlearned, while $\mathcal{Q}_h$ represents the distribution of models where $\{z\}$ was never part of the training set and thus not unlearned. Hence, given an unlearned model $\theta_\mathcal{U}$, the attacker makes a likelihood ratio test~\cite{carlini_lira} to derive the membership of the target sample according to \autoref{eq:test_privacy}. 

\vspace{-0.1in}

\begin{align}
\Lambda(z) = \frac{p(\theta_{\mathcal{U}} \mid Q_{{u}}(z))}{p(\theta_{\mathcal{U}} \mid Q_{\text{h}}(z))}, \quad \forall (z) \in D_{\text{target}}
\label{eq:test_privacy}
\end{align}

Furthermore, regarding \textbf{Game~\ref{game3}}, attacker essentially requires these distributions:

\vspace{-0.1in}

\begin{align*}
\mathcal{Q}_u =& \{\theta \leftarrow \mathcal{U}(\mathcal{A}(D_{\text{train}} \cup \{z\}), \{z\} \cup D_f \mid D_f\subset D_{\text{train}} \sim \mathcal{D})\} \\
\mathcal{Q}_R = & \{\theta \leftarrow \mathcal{A}(D_{\text{train}} \mid D_{\text{train}} \sim \mathcal{D})\}
\end{align*}

Similarly, $\mathcal{Q}_R$ represents the distribution of the models trained without the target sample ($z$) (``\emph{retrained excluding $\{z\}$}''). Different from previous tests, the hypothesis test here cannot be on a single unlearned model or retrained model. Hence, we introduce a new \textit{unlearning test model} as: 

\vspace{-0.1in}

\begin{align}
\theta_{\mathcal{T}}(z) = 
\begin{cases} 
\theta_\mathcal{U}(z) & \text{if } z \in D_\text{target} \cap D_\text{train} \\
\theta_{{\mathcal{R}}}(z) & \text{if } z \in D_\text{target} \setminus D_\text{train}
\end{cases}
\label{eq:test_function}
\end{align}

Then, given a test model $\theta_\mathcal{T}$, the following likelihood ratio test can be adopted to derive the efficacy based on \textbf{Game~\ref{game3}}. 

\vspace{-0.1in}

\begin{align}
\Psi(z) = \frac{p(\theta_{\mathcal{T}} \mid Q_{{u}}(z))}{p(\theta_{\mathcal{T}} \mid Q_{{R}}(z))}, \quad \forall (z) \in D_{\text{target}}
\label{eq:test_efficacy}
\end{align}

In practice, the test model is introduced to enable per-sample efficacy evaluation without requiring direct access to both the unlearned and retrained models for comparison. Since each target sample may either be unlearned or never trained on, but we only have access to individual models trained under one configuration at a time, we use $\theta_{\mathcal{T}}(z)$ as a unified interface. It returns the output from the correct model depending on $\{z\}$'s status: unlearned or excluded. \autoref{fig:revision_framework} (b) shows how a test model operates in \sys. If a target sample is unlearned, a query would be given from the unlearned model. Otherwise, the query would be given from the model that is trained, excluding the target sample, and the efficacy would be evaluated from a likelihood test per \autoref{eq:test_efficacy}.

\subsubsection{MIA in \sys}
\label{MIA_subsection}

\noindent\textbf{Membership Inference in \sys.}  
Algorithm~\ref{alg:lira} outlines the attack pipeline in \sys to evaluate the privacy leakage and unlearning efficacy. The attacker trains $N$ shadow models using samples drawn from the data distribution $\mathcal{D}$. For each target sample $z$, the algorithm ensures that model outputs (observations) are collected under three training scenarios: when $z$ is included in training (\textit{In}), when it is excluded entirely (\textit{Out}), and when it is included and later unlearned via a known unlearning algorithm (\textit{Unlearned}). It also simulates when it is \textit{Out} and the model is unlearned  (\textit{Held-out}). After each model is trained, the inference function $\phi$ is applied to obtain the model's observation on the target sample, and the result is stored in one of five observation sets. 
\autoref{fig:revision_framework} shows an example for the shadow model training and unlearning.

In the second step, Kernel Density Estimation (KDE)~\cite{silverman1986density} is applied to fit smooth distributions over the collected confidence scores from each distribution. The attacker then queries the actual unlearned model and computes two likelihood ratio tests: $\Lambda$, which compares the unlearned distribution to the held-out distribution to evaluate \emph{privacy leakage}, and $\Psi$, which compares the unlearned distribution to the out distribution (excluded samples) to measure the unlearning \emph{efficacy}. 

\vspace{0.05in}

\noindent \textbf{Parallelizing MIA to All Target Samples}. To scale the algorithm to a set of target samples while maintaining balanced per-sample coverage across roles, we structure the shadow training in groups of three. In each iteration, we select a disjoint subset of target samples of size $\frac{N}{3}$, and assign each sample to a unique role within that batch: one-third as $D_\text{in}$, one-third as ${D_\text{unlearn}}$, and one-third as $D_\text{out}$. We train shadow models $D_\text{in}\cup D_\text{unlearn}\cup D_\text{attack}$ and then unlearn $D_\text{unlearn}$. After repeating this procedure three times, each target sample will have at least one observation instance for all intended distributions. Once all $N$ iterations are completed, this structure guarantees that each target sample has at least $\frac{N}{3}$ complete set of observations to form the corresponding distributions ($\mathcal{Q}_u, \mathcal{Q}_R, \mathcal{Q}_h$). The resulting distributions are then used to fit KDE distributions for the privacy leakage and efficacy evaluations of the target samples.

\input{algorithms/MIA}

\vspace{0.05in}

\noindent\textbf{RULI vs. U-LiRA~\cite{hayes2024inexact}}. \label{ulira-limit} One might naturally extend the MIAs to the unlearning setting by considering the unlearning step, comparing the distributions of retrained and unlearned models with a query from an unlearned model and formulating a likelihood-based ratio test such as:
\[
\frac{p(\theta_{\mathcal{U}} \mid \mathcal{Q}_{u}(z))}{p(\theta_{\mathcal{U}} \mid \mathcal{Q}_{R}(z))}.
\]
This is the core idea behind {U-LiRA}~\cite{hayes2024inexact}, the first per-sample framework for evaluating unlearning by comparing inferences from the unlearned model ($\mathcal{Q}_u$) against those of a retrained model ($\mathcal{Q}_R$). U-LiRA marks a notable step towards structured evaluation of unlearning effectiveness. Then, we summarize the major differences between \sys and U-LiRA as below. 

\vspace{0.5em}

    
    \noindent\emph{Efficacy Modeling}. The hypothesis test in U-LiRA (as discussed above) relies on a critical assumption: the inference behavior of the unlearned model, $\theta_{\mathcal{U}}(z)$, will closely match that of the retrained model, $\theta_{R}(z)$, for target samples $z \in D_{\text{target}} \setminus D_{\text{train}}$. Membership is then inferred by comparing the unlearned model's output at $z$ to the distribution $\mathcal{Q}_R(z)$. 
    
    However, this assumption may not hold. Our empirical analysis (see~\autoref{Q_U_LIRA}) reveals a distributional mismatch: the unlearned and retrained models diverge significantly where $z$ is out. This divergence is not merely a subtle shift—it directly impacts MIA performance and may lead to impacting privacy leakage and more noticeably efficacy measurement. Also, this mismatch reflects a limitation of Game~\autoref{game2} and our motivation for introducing Game~\autoref{game3} and consequently the Test model for efficacy measurement in \sys.  

\vspace{0.05in}
    
    \noindent \emph{Targeted Attack for Unlearning Evaluation}. First, in contrast to U-LiRA, \sys incorporates an additional shadow distribution $\mathcal{Q}_h$, which represents held-out samples, i.e., samples that are neither trained nor unlearned. This enables a dual-objective inference framework where privacy leakage $\Lambda$ is evaluated against $\mathcal{Q}_h$, and efficacy $\Psi$ is evaluated against $\mathcal{Q}_R$ (not in U-LiRA). Second, this requires a revised shadow model design, and \sys remains computationally efficient by sharing models across roles. Third, \sys involves a new target sample selection strategy on vulnerable samples (canaries) while U-LiRA focuses on random samples. We demonstrate that applying our target selection strategy to U-LiRA improves its MIA performance, outperforming previously reported accuracy even for strong unlearning baselines. Finally, as a targeted attack, \sys requires significantly fewer shadow models, making it more efficient than U-LiRA. We provide empirical comparisons for \sys and U-LiRA in Section~\ref{vs_ulira-results}.

\section{Rectified Unlearning Evaluation}
\label{sec:unlearning_eva}

In this section, we integrate the proposed MIA in \sys into a unified framework for unlearning evaluation. Section~\ref{steps_1} illustrates how these components can interconnect to provide a holistic view of privacy risks and efficacy. With the refined likelihood ratios, we then perform targeted attacks on canaries using the selection strategy from Section~\ref{sec:canary}.

\subsection{Major Steps for Rectified Evaluation}
\label{steps_1}

\noindent\textbf{Shadow Model Training.}  
Using the MIA in \sys, we train $N$ shadow models with a fixed target unlearning set. For the shadow-trained models, we obtain the \textit{In} and \textit{Out} distributions. 
For the shadow-unlearned models, each sample $x \in D_\text{target}$ is evenly distributed across three inference types: \textit{remained}, \textit{unlearned}, and \textit{held-out}, with each type represented by $\frac{N}{3}$ models. This corresponds to Steps 1 and 2 in Algorithm~\autoref{alg:lira}.

\vspace{0.05in}

\noindent\textbf{Target Model Training and Unlearning.} 
We begin by partitioning the unlearning target set $D_\text{target}$ into three equal subsets: one-third is designated to be excluded from training, while the remaining two-thirds are used to train the target model $\theta_\mathcal{I}$ along with disjoint attack data to ensure generalization. After training, we perform unlearning by removing half of the $D_\text{target}$ training subset, resulting in the unlearned model $\theta_\mathcal{U}$. This setup ensures a balanced design for evaluating both validation accuracy and membership inference (MI) using the two likelihood tests defined in \sys: $\Lambda$ (privacy leakage) and $\Psi$ (unlearning efficacy), as detailed in Steps 3 and 4 in Algorithm~\autoref{alg:lira}. Since these evaluations require equal-sized subsets for fair comparison, we may adjust the partitioning ratio of $D_\text{target}$ accordingly if needed.

Note that \sys can provide a comprehensive set of likelihood tests in ~\autoref{tab:ratio_test} with a single run. It enables measurement metrics other than privacy leakage and efficacy on forget data. For example, we might evaluate how unlearning impacts remain data privacy and how memorization levels would be changed. Ideally, the memorization on remain samples should remain similar to the original model (no unintended privacy leakage and no unintended unlearning).  

\vspace{-0.05in}

\input{Tables/notations}

\subsection{Target Samples Selection}
\label{sec:canary}

To bridge the proposed attack with the evaluation, we now address a critical challenge: How to select target samples that represent corner cases in unlearning? Most existing unlearning methods rely on random subsets of the training data, focusing on average-case evaluations. However, not all samples are memorized equally, and most are generally well-protected in typical datasets~\cite{gradients_looklike}. This inspires us to investigate unlearning in a more challenging scenario: How does unlearning perform on highly memorized (vulnerable) samples? 

\begin{figure}[!h]
    \centering    \includegraphics[width=0.9\linewidth]{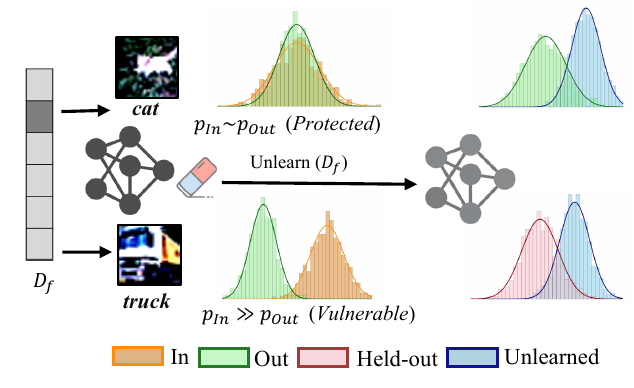}
    \vspace{-0.1in}
    \caption{\text Two samples under inexact unlearning; A protected sample's out and unlearned distributions are distinguishable (efficacy); vulnerable sample's unlearned and held-out distributions are distinguishable (privacy leakage).}\vspace{-0.1in} 
    \label{fig:case_vulnerable}
\end{figure}

We further examine the impact of unlearning vulnerable samples alongside non-vulnerable (i.e., safe or protected) ones, inspired by the canary injection technique from privacy auditing~\cite{one_line, nasr2023tight, carlini2019secret}. Our results show that, when carefully tuned, stronger unlearning methods can provide greater protection than expected—especially \textit{when only vulnerable samples are unlearned} (see~\autoref{tab:scrub_NegGrad} and~\autoref{table:vit_experiments}).  
We found that a more practical setting of injecting vulnerable samples as canaries yields higher privacy leakage and lower efficacy (higher MIA success). We suspect this behavior is influenced by the relative impact of sample vulnerability, and when samples with different memorization levels are unlearned in mini-batches, the averaged gradient cannot update the model to sufficiently unlearn the vulnerable samples. An example of such an effect in~\autoref{fig:case_vulnerable} where unlearning vulnerable samples together with protected samples is challenging. We leave a deeper investigation of this observation to future work. For now, we consider this scenario the most challenging and practical setting for unlearning. To support this, we have conducted a comprehensive study on alternative target sample selection strategies in Section~\ref{sec:diff_target}.

\vspace{-0.2in}

\noindent 

\noindent \textbf{Canary Injection.} 
By specifying the target set in \sys, we can focus on measuring privacy leakage (rather than efficacy), drawing inspiration from the well-known canary injection technique~\cite{one_line, carlini2019secret} used in differential privacy auditing~\cite{nasr2023tight}. This approach offers an efficient strategy for evaluating inexact unlearning algorithms on their ability to remove vulnerable samples. The strategy is simple: inject canaries into the forget set and query only these samples after unlearning.

Specifically, we inject vulnerable samples into a randomly selected forget set. This random forget set may or may not overlap with a predefined set of \textit{safe} samples, allowing us to simulate realistic unlearning scenarios (a portion of forget data can be vulnerable). Then, we only focus on the attacker's advantage solely on vulnerable samples. However, one might try to achieve tighter bounds on the privacy leakage of unlearning by a more careful choice of canaries or adversarial canary injection. In this paper, we focus on providing more insights for unlearning evaluation rather than trying to achieve very tight bounds on the privacy leakage of unlearned models.

%% file: algorithms/MIA.tex
\setcounter{algorithm}{0}
\begin{algorithm}[!ht]
\small
\caption{Membership Inference Attack in \sys} 
\label{alg:lira}
\begin{algorithmic}[1]
\Require trained model $f$, target sample $z \in D_\text{target}$, forget data $D_f \in D_{\text{attack}}$, data distribution $\mathcal{D}$, training algorithm $\mathcal{A}$, unlearning algorithm $\mathcal{U}$, inference function $\phi$ (loss, logit-scaled confidence), output observation $\mathcal{O}$, number of shadow models $N$.

\State \textbf{Initialize:} $\textbf{$\mathcal{O}$}_{\text{in}}, \textbf{$\mathcal{O}$}_{\text{out}}, \textbf{$\mathcal{O}$}_{\text{unlearned}}, \textbf{$\mathcal{O}$}_{\text{held-out}}$
\State $\textbf{$\mathcal{O}$}_{\text{remained}} \gets \emptyset$


\Statex \hspace*{-\algorithmicindent} \textcolor{blue}{\texttt{▸ Step 1: training \& unlearning shadow models}}

\For{$N$ iterations}

    \State $D_{\text{attack}} \leftarrow \mathcal{D}$
    
    \State $f_i \leftarrow \mathcal{A}(D_{\text{attack}} \cup \{z\})$
    \State $f_R \leftarrow \mathcal{A}(D_{\text{attack}})$

    \State $\textbf{$\mathcal{O}$}_{\text{in}} \gets \textbf{$\mathcal{O}$}_{\text{in}} \cup \{\phi(f_i(z))\}$ 
    \State $\textbf{$\mathcal{O}$}_{\text{out}} \gets \textbf{$\mathcal{O}$}_{\text{out}} \cup \{\phi(f_R(z))\}$ 
    
    \State $f_{\text{u}} \leftarrow \mathcal{U}(\mathcal{A}(D_{\text{attack}} \cup \{z\}), D_f \cup \{z\})$
    \State $\textbf{$\mathcal{O}$}_{\text{unlearned}} \gets \textbf{$\mathcal{O}$}_{\text{unlearned}} \cup \{\phi(f_{\text{u}}(z))\}$

    \State $f_{\text{h}} \leftarrow \mathcal{U}(\mathcal{A}(D_{\text{attack}}), D_f)$
    \State $\textbf{$\mathcal{O}$}_{\text{held-out}} \gets \textbf{$\mathcal{O}$}_{\text{held-out}} \cup \{\phi(f_{\text{h}}(z))\}$

    \State $f_{\text{r}} \leftarrow \mathcal{U}(\mathcal{A}(D_{\text{attack}} \cup \{z\}), D_f)$
    \State $\textbf{$\mathcal{O}$}_{\text{remained}} \gets \textbf{$\mathcal{O}$}_{\text{remained}} \cup \{\phi(f_{\text{r}}(z))\}$ 
\EndFor

\Statex \hspace*{-\algorithmicindent} \textcolor{blue}{\texttt{▸ Step 2: estimating distributions}}

\State $\hat{f}_{\text{in}} \gets \mathtt{KDE}(\textbf{$\mathcal{O}$}_{\text{in}})$
\State $\hat{f}_{\text{out}} \gets \mathtt{KDE}(\textbf{$\mathcal{O}$}_{\text{out}})$
\State $\hat{f}_{\text{unlearned}} \gets \mathtt{KDE}(\textbf{$\mathcal{O}$}_{\text{unlearned}})$
\State $\hat{f}_{\text{held-out}} \gets \mathtt{KDE}(\textbf{$\mathcal{O}$}_{\text{held-out}})$
\State $\hat{f}_{\text{remained}} \gets \mathtt{KDE}(\textbf{$\mathcal{O}$}_{\text{remained}})$

\Statex \hspace*{-\algorithmicindent} \textcolor{blue}{\texttt{▸ Step 3: querying models}}


\State $f_U \gets \mathcal{U}(f, {z})$ 
\State $\textbf{$\mathcal{O}$}_{f_U} \gets \phi(f_{\text{U}}(z))$

\State $f_T \gets$ \autoref{eq:test_function}
\State $\textbf{$\mathcal{O}$}_{f_T} \gets \phi(f_{\text{U}}(z))$

\Statex \hspace*{-\algorithmicindent} \textcolor{blue}{\texttt{▸ Step 4: deriving the privacy leakage \& efficacy}}


\State $\Lambda \gets \frac{p(\textbf{$\mathcal{O}$}_{f_U} \mid \hat{f}_{\text{unlearned}})}{p(\textbf{$\mathcal{O}$}_{f_U} \mid \hat{f}_{\text{held-out}})}$
\State $\Psi \gets \frac{p(\textbf{$\mathcal{O}$}_{f_T} \mid \hat{f}_{\text{unlearned}})}{p(\textbf{$\mathcal{O}$}_{f_T} \mid \hat{f}_{\text{out}})}$

\State \textbf{Return} $\Lambda$, $\Psi$
\end{algorithmic}
\end{algorithm}

%% file: Tables/notations.tex
\begin{table}[!h]
\small
\centering
\footnotesize
\caption{\sys enables comprehensive MIA tests on unlearned and trained models.}
\vspace{-0.1in}
\resizebox{\columnwidth}{!}{
\begin{tabular}{@{}l|c@{}}
\toprule
Target Evaluation & Likelihood Ratio Test \\ \midrule
Unlearning efficacy & $\frac{p(\theta_{\mathcal{T}} \mid \mathcal{Q}_{{u}}(z))}{p(\theta_{\mathcal{T}} \mid \mathcal{Q}_{{R}}(z))}, \quad \forall (z) \in D_{\text{target}}$ \\
Privacy leakage & $\frac{p(\theta_{\mathcal{U}} \mid \mathcal{Q}_{{u}}(z))}{p(\theta_{\mathcal{U}} \mid \mathcal{Q}_{{h}}(z))}, \quad \forall (z) \in D_{\text{target}}$ \\
Trained model privacy leakage & $\frac{p(\theta_{\mathcal{I}} \mid \mathcal{Q}_{{i}}(z))}{p(\theta_{\mathcal{I}} \mid \mathcal{Q}_{{R}}(z))}, \quad \forall (z) \in D_{\text{target}}$ \\
Privacy leakage on remained samples  & $\frac{p(\theta_{\mathcal{U}} \mid \mathcal{Q}_{{r}}(z))}{p(\theta_{\mathcal{U}} \mid \mathcal{Q}_{{h}}(z))}, \quad \forall (z) \in D_{\text{target}}$ \\
\bottomrule
\end{tabular}
}\vspace{-0.15in}
\label{tab:ratio_test}
\end{table}

%% file: Evaluation.tex
\section{Experiments}
\label{sec:exp}

\input{evaluation/setup}

\input{evaluation/evaluation_efficacy}

\subsection{Comparison with U-LiRA}
\label{vs_ulira-results}

\noindent\textbf{Differences in Target Design and Testing Strategy.}  
\sys differs from U-LiRA along two key designs: target sample selection and model querying during hypothesis testing. U-LiRA selects random samples from a single class (the setting of ``Class''), while we construct the target set by injecting vulnerable samples into a set of known protected samples, though we only evaluate on the vulnerable ones. Additionally, while both U-LiRA and \sys query the unlearned model \(\theta_\mathcal{U}\) for privacy leakage, they differ in how reference distributions are constructed. U-LiRA compares outputs between unlearned and retrained models (\(\mathcal{Q}_u\) vs \(\mathcal{Q}_R\)), while \sys uses \(\mathcal{Q}_h\), a held-out distribution \(\mathcal{Q}_R\). For efficacy, \sys evaluates queries using the test model \(\theta_\mathcal{T}(z)\), which selects between a retrained model \(\theta_\mathcal{R}\) or an unlearned model \(\theta_\mathcal{U}\) depending on whether the sample was excluded or unlearned.

\input{Tables/revision_comparison}

\vspace{0.05in}

\noindent\textbf{Results}.\footnote{We re-implemented U-LiRA (not official) based on the algorithm in~\cite{hayes2024inexact}. 
}~\autoref{tab:revision_compare_ulira} evaluates the impact of two key components in our evaluation pipeline: (1) the target sample selection strategy, and (2) the MIA. We compare four configurations using both $\ell_1$ Sparse and Scrub unlearning (\emph{which achieve the top unlearning performance from prior results}). U-LiRA's original setting selects random samples from ``Class'' as the target set and uses their own likelihood-ratio test. In contrast, our setting structures the target by injecting vulnerable samples into a set of protected samples, and uses \sys for the attack.

When we replace only the target selection while keeping U-LiRA's test, privacy leakage improves significantly. Under Scrub, TPR@1\%FPR increases from 1.13\% to 8.6\%. This shows that our target selection alone improves the evaluation, even benefiting U-LiRA. Conversely, when we change only the attack to \sys while keeping the ``Class'' target, the gains are marginal, suggesting that these targets limit the attack performance. Finally, in our full setting (Vulnerable + Protected target + \sys), we observe the strongest results, e.g., Scrub achieves 11.8\% TPR@1\%FPR for privacy leakage and 8.8\% for efficacy, while $\ell_1$ Sparse reaches 1.4\% and 23.3\%, respectively. Since U-LiRA assesses efficacy through the lens of privacy leakage, we do not report MIA results for its efficacy. 

\vspace{0.05in}

\noindent\textbf{Efficiency}. We acknowledge that training shadow models introduces additional computational costs, and this applies to all per-sample standard MIA evaluations. For reference, \sys incurs ~1.2$\times$ runtime of LiRA~\cite{carlini_lira} (standard per-sample MIA for machine learning training) per shadow model under unlearning CIFAR-10 using Scrub. In practice, U-LiRA is more computationally expensive. For example, in a setting with 1,200 target samples (e.g., our Vulnerable + Protected setting), for preparing 30 unlearned shadow models per sample, U-LiRA incurs approximately 270$\times$ higher cost than single shadow model training and requires around 40$\times$ unlearning models for each trained model. While, \sys completes it with 36$\times$ more than single shadow model training.

\subsection{Generalizability of \sys}

We next evaluate the generalizability of \sys by applying it to model fine-tuning on larger-scale and complex datasets (TinyImageNet and WikiText-103) and unlearning in vision transformer (ViT) and language models.

\subsubsection{Vision Transformer (ViT) on TinyImageNet}

For the TinyImageNet dataset, we employ a pre-trained ViT (Swin-small~\cite{liu2021swin} Transformer) to fine-tune and unlearn. Training is performed to reach the test accuracy of at least 84\% by fine-tuning the model only with 2 epochs. Shadow model training and canary injection settings are similar to those on preparing 90 shadow models in Sections \ref{mia_setting} and \ref{canary_setting}. For unlearning evaluation, 
we avoid from-scratch training or fine-tuning architectures like ResNets on this dataset, due to low test accuracy and severe overfitting, making them unsuitable for our study. We also use the top benchmarks from prior results ($\ell_1$ Sparse and Scrub) to unlearn image samples from the ViT model for this group of experiments.

\begin{figure}[h]
    \centering
    \begin{subfigure}[b]{0.5\columnwidth}
        \centering
        \includegraphics[width=\textwidth]{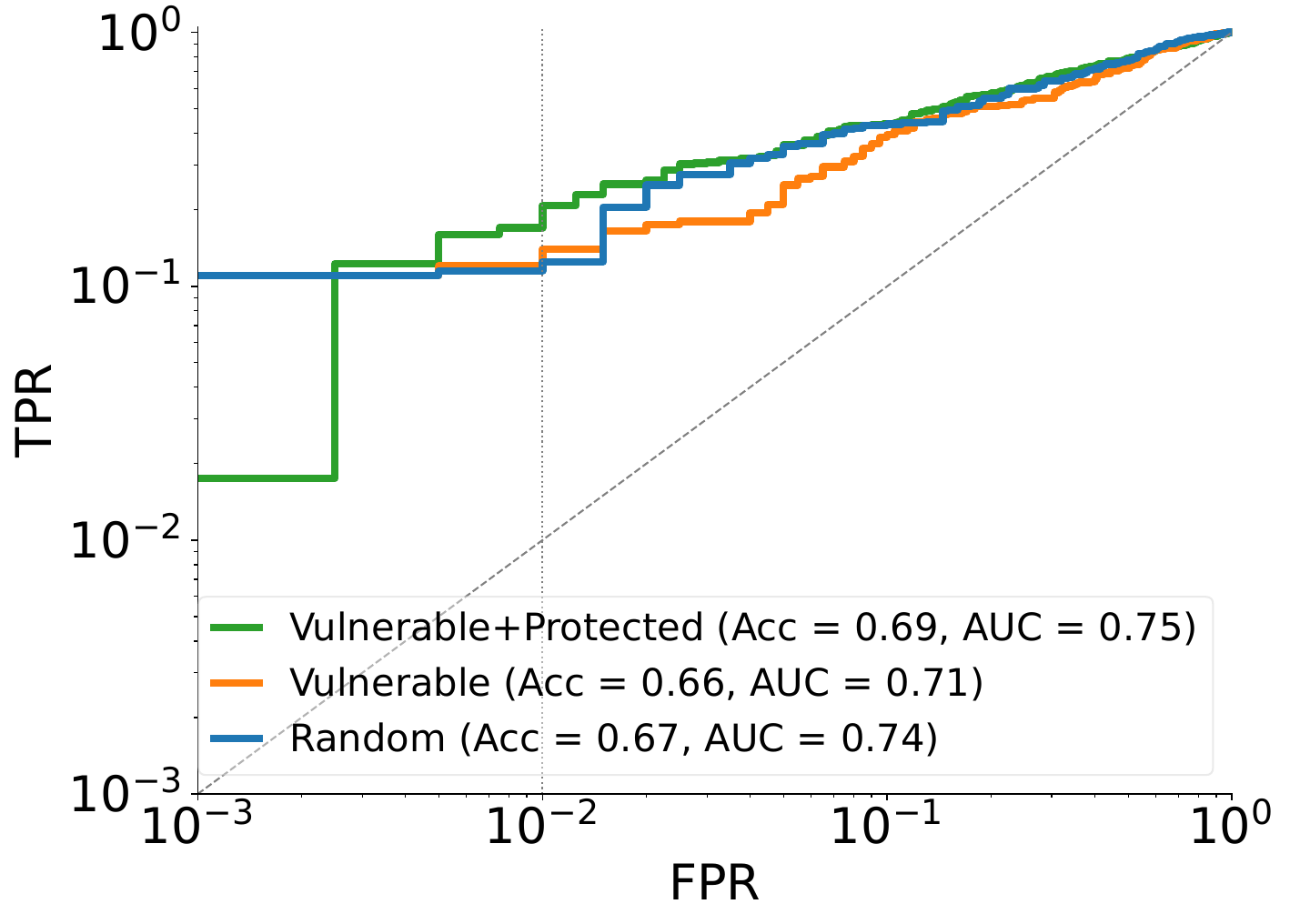}
        \caption{$\ell_1$ Sparse}
        \label{fig:sparse_roc_revision}
    \end{subfigure}
    \hspace{-0.08in}
    \begin{subfigure}[b]{0.5\columnwidth}
        \centering
        \includegraphics[width=\textwidth]{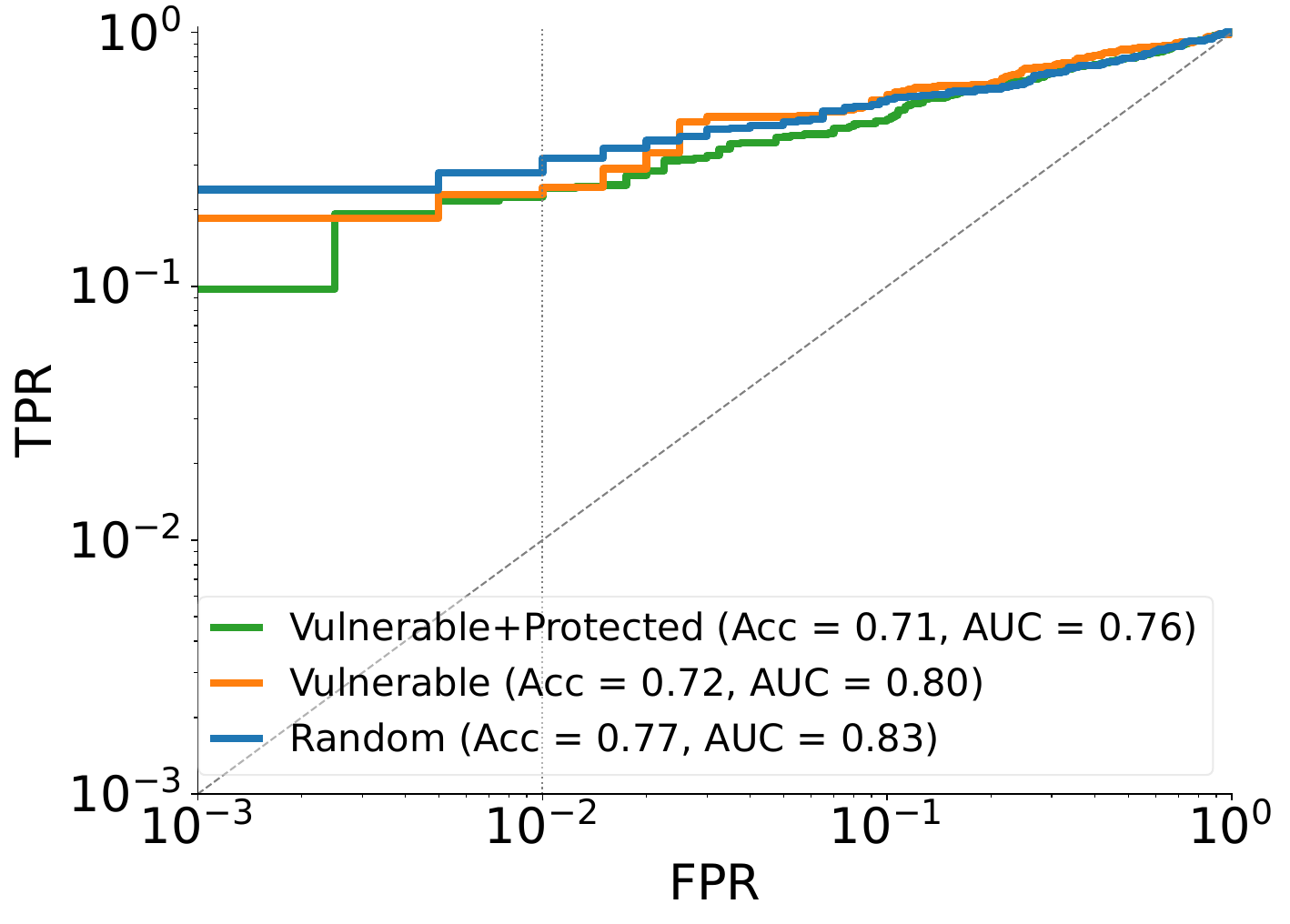}
        \caption{Scrub}
        \label{fig:scrub_roc_revision}
    \end{subfigure}
    \vspace{-0.15in}
    \caption{MIA for unlearning efficacy on different choices of target samples, with best unlearning baselines evaluated when fine-tuning and unlearning a Swin-small ViT model.} 
    \label{fig:efficacy_vit}
\end{figure}

\input{Tables/vit_experiments}

\vspace{0.05in}

\noindent\textbf{Results.} Efficacy results indicate a consistent gap in per-sample behavior between the retrained model and unlearned models across all three target data configurations, achieving an attack accuracy of at least 66\% on $\ell_1$ Sparse and 71\% on Scrub. Efficacy tests can achieve at least 10\% TPR@1\%FPR in the ``Vulnerable + Protected'' setting. The privacy leakage results show a different trend. We conduct a targeted population attack as the baseline and compare it with \sys on a different set of target data. Results confirm the prior observations: unlearning vulnerable samples, when injected as canaries, leads to increased privacy leakage. Exclusively unlearning vulnerable samples results in a higher attack success rate compared to random samples, but is not generally more effective than when they are injected as canaries. 

\subsubsection{Text Generation on WikiText-103}
We conduct our experiments on the WikiText-103 dataset, fine-tuning Pythia-70m ~\cite{biderman2023pythia} and GPT2-small~\cite{radford2019language} (see details in Appendix~\ref{further_implementation_detailss}). The unlearning task focuses on removing \textit{n-gram sequences} that the fine-tuned models have memorized within the training. To validate \sys's generalizability, we apply unlearning methods designed for language models~\cite{zhang2024negative-NPO, jang-etal-2023-knowledge, liu2022continual-GALLM, shi2024muse}. The attacker aims to infer the membership of the specific n-gram sequence by querying the unlearned model.

\vspace{0.05in}

\noindent\textbf{Unlearning Benchmarks.} 
Previous inexact unlearning methods in Section~\ref{unleanring_benchmarks}, are not originally proposed for language model unlearning. Instead, we adapt two gradient-based unlearning methods~\cite{shi2024muse}: {GA+GDR} and {NPO~\cite{zhang2024negative-NPO}} to unlearn sequences from language models. GA+GDR applies Gradient Ascent on the forget data to increase its loss and reduce its influence, while using Gradient Descent on the remain data (GDR) to preserve generalization. NPO introduces a directional forgetting loss that compares the current model's logits to those of a frozen trained model (\textit{reference model}), directing the updated model to assign lower confidence to the forget data. In both methods, the forgetting and retention objectives are applied jointly to ensure effective removal without degrading overall performance. Moreover, we add SFT steps on remaining data to recover any loss due to unlearning.  

\vspace{0.05in}

\noindent \textbf{Attack Setup.} For shadow model training, we use a dataset of 15,000 records (containing over 1 million tokens) and consider 1,000 7-gram sequences~\cite{duan2024membership} (with prefixes drawn from test data records) as the target data for training shadow models. We fine-tune both Pythia-70m and GPT2-small with 5 supervised fine-tuning (SFT)~\cite{brown2020language_sft} epochs and 2 epochs of prefix language modeling (PLM)~\cite{raffel2020exploring_prefixlanguagemodeling}. Similarly, for unlearning, we use PLM to optimize the model to unlearn the target 7-grams. We also execute 2 more SFT epochs on the remaining data to improve unlearning and generalization (see more details in Appendix~\ref{further_implementation_detailss}). We have trained 60 shadow models and collected the target 7-grams samples' loss~\cite{yeom_mia_loss, panda2025privacy} for membership signaling. 
Note that canary injection was not applied, as it is not yet commonly adopted in language models; nonetheless, \sys still demonstrates strong performance in evaluating text unlearning, as detailed below. 

\vspace{-0.05in}

\begin{figure}[h]
    \centering
        \begin{subfigure}[t]{0.5\linewidth}
            \centering
            \includegraphics[width=\linewidth]{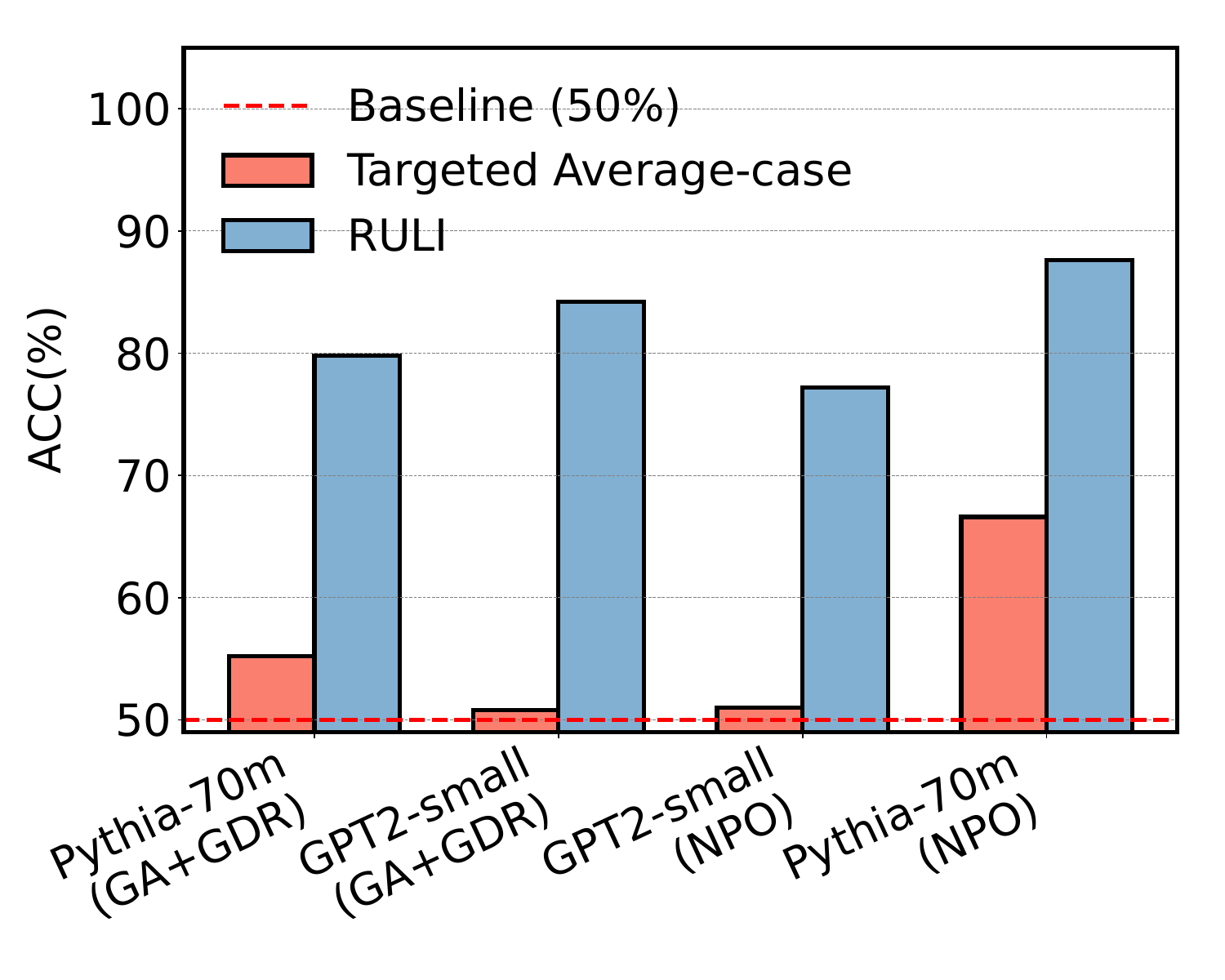}
        \caption{Attack accuracy results}
        \end{subfigure}
        \hspace{-0.1in}
        \begin{subfigure}[t]{0.5\linewidth}
            \centering
            \includegraphics[width=\linewidth]{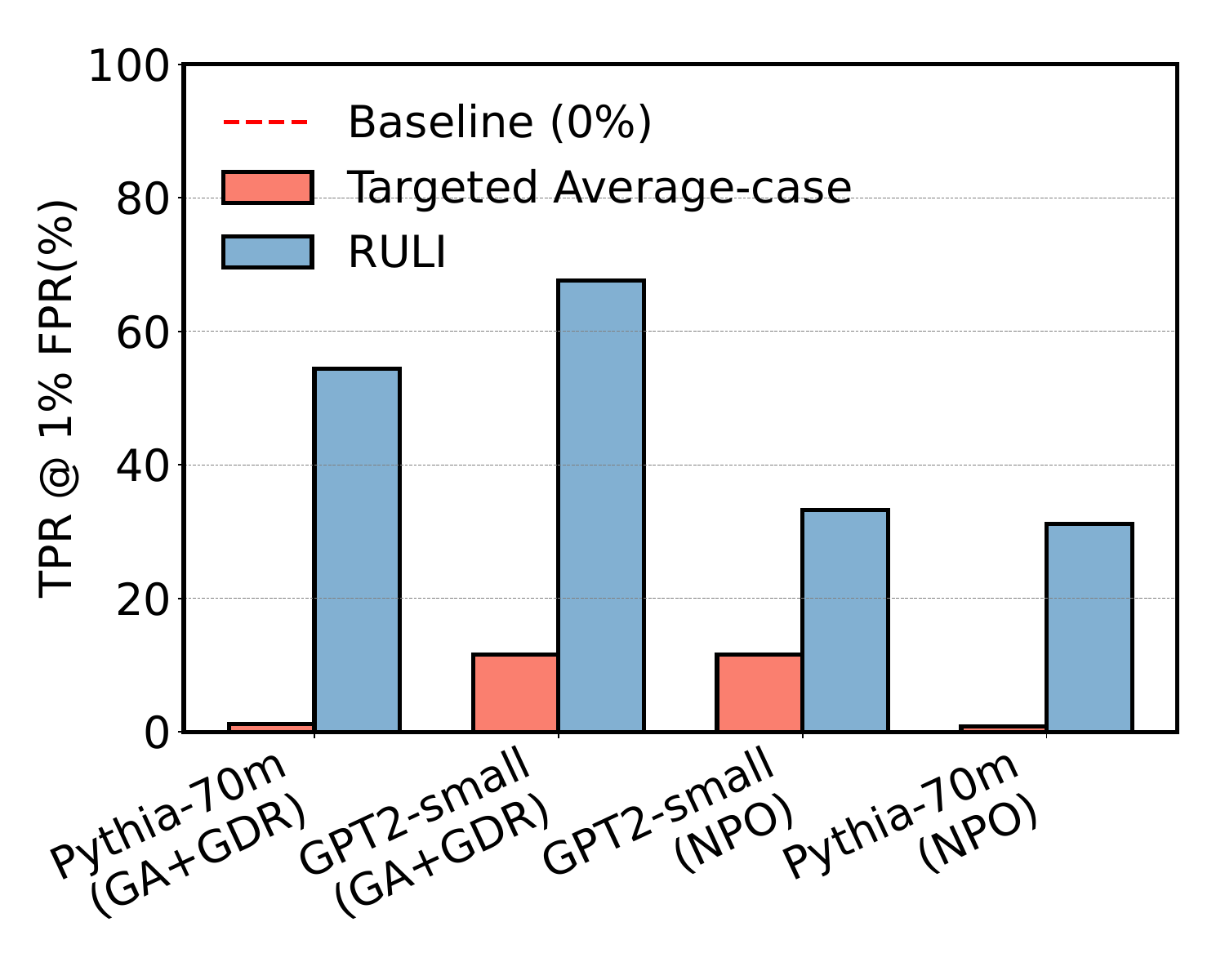}
            \caption{TPR@1\%FPR results}
        \end{subfigure}\vspace{-0.1in}
        \caption{Privacy leakage results on WikiText-103 using a target set of 500 samples (\textit{250 out and 250 unlearned}) on WikiText-103 with the unlearning 7-gram sequences task.}\vspace{-0.1in}
        \label{fig:LLM_results_leakage}
\end{figure}

\begin{figure}[h]
    \centering
        \begin{subfigure}[t]{0.5\linewidth}
            \centering
            \includegraphics[width=\linewidth]{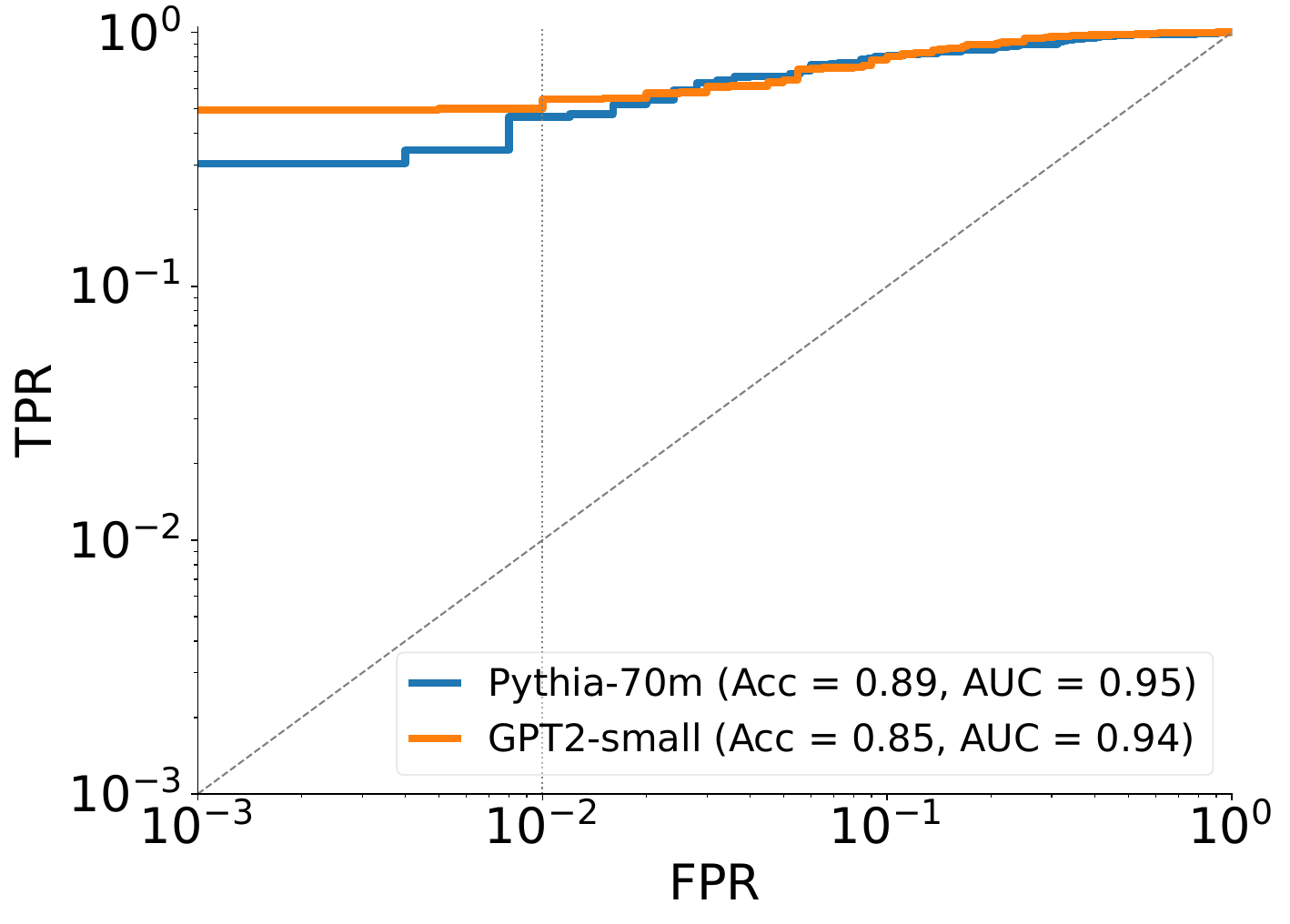}
        \caption{Attack accuracy results}
        \end{subfigure}
        \hspace{-0.08in}
        \begin{subfigure}[t]{0.5\linewidth}
            \centering
            \includegraphics[width=\linewidth]{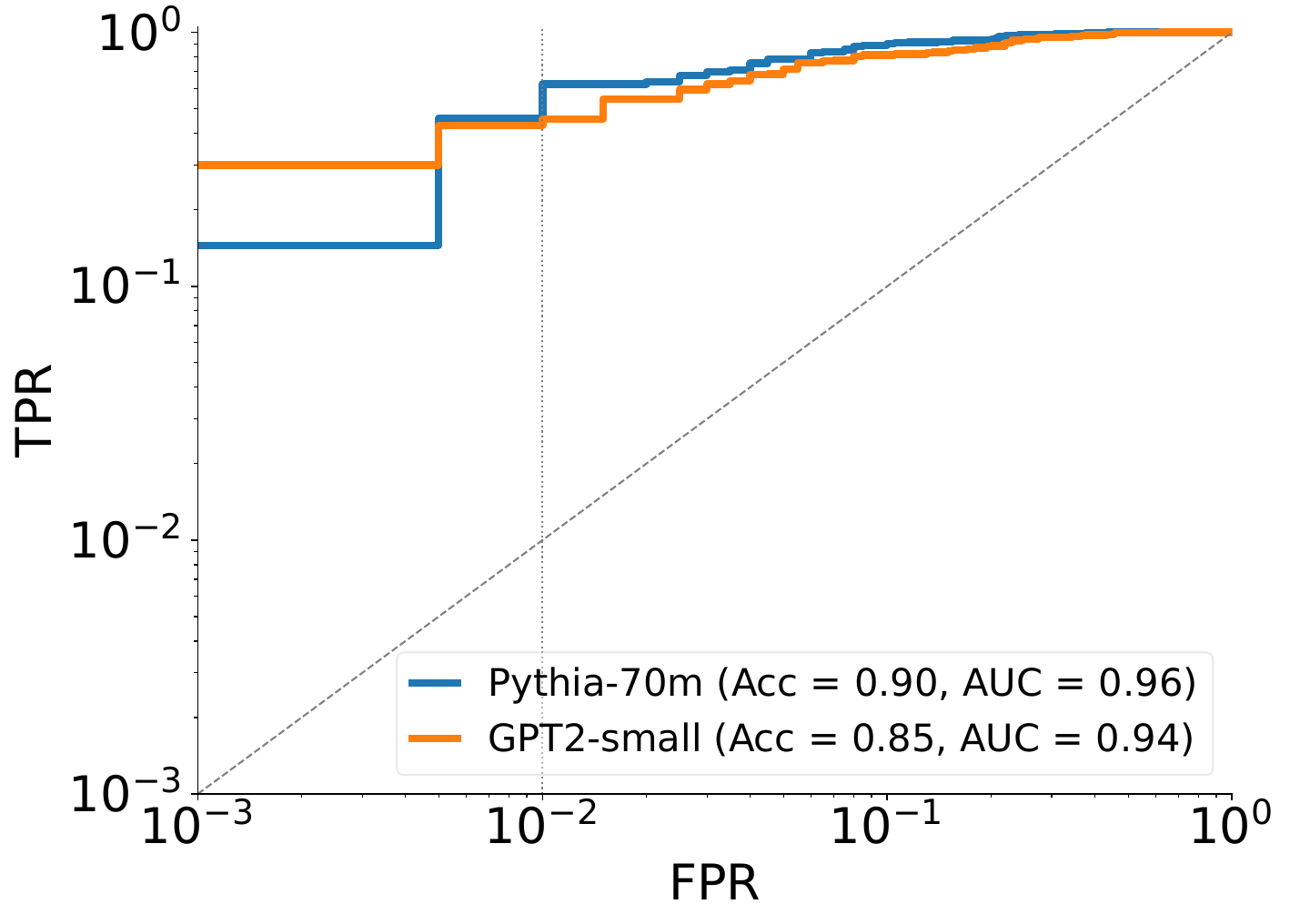}
            \caption{TPR@1\%FPR results}
        \end{subfigure}\vspace{-0.05in}
        \caption{Unlearning efficacy on 400 target samples (\textit{200 Out and 200 Unlearned}), evaluated using NPO and GA+GDR baselines with fine-tuning-based unlearning on GPT2-small and Pythia-70m models.}\vspace{-0.1in}
        \label{fig:LLM_results_efficacy}
\end{figure}

\vspace{0.05in}

\noindent \textbf{Results.} 
\autoref{fig:LLM_results_leakage} shows that \sys achieves higher accuracy and TPR@1\%FPR than the targeted average-case baseline across all evaluated models and unlearning methods. The Pythia-70m model generally yields more privacy leakage than GPT2-small, especially under the GA+GDR unlearning method. \autoref{fig:LLM_results_efficacy} shows that both GA+GDR and NPO exhibit substantial differences from the retrained model, with Pythia-70m reaching up to 90\% MIA accuracy. This suggests that the unlearned model still retains distinguishable characteristics compared to the retrained model. 

\vspace{0.05in}

\noindent\textbf{Scaling to LLMs.} While we demonstrate the effectiveness of \sys in multiple domains, including TinyImageNet and WikiText-103, there remain several challenges in applying it to production-scale language models with broad knowledge coverage. First, performing per-sample MIAs on LLMs is computationally intensive and may be impractical at scale. Second, privacy and unlearning on text data and LLMs have not yet been well standardized~\cite{shumailov2024ununlearning}, and current membership-based auditing methods may not fully capture privacy leakage in LLMs~\cite{mireshghallah2024can}. Thus, while \sys is applicable to relatively smaller-scale settings for unlearning text in language models~\cite{panda2025privacy}, privacy and efficacy evaluation for LLM unlearning needs further advancements, along with the future development of unlearning guarantees and privacy attacks on LLMs.

%% file: evaluation/setup.tex
\subsection{Experimental Setup}
Consistent with existing research on unlearning \cite{thudi2022unrolling, scrub, golatkar2020eternal,sparsity,hayes2024inexact} and machine learning privacy benchmarks~\cite{carlini_lira,hayes2024inexact,population_1}, we evaluated the unlearning methods on typical CIFAR-10 and CIFAR-100 datasets with ResNet models. To demonstrate the generalizability of \sys, we evaluate unlearning on two challenging settings: TinyImageNet fine-tuned on a vision transformer (ViT), and WikiText-103 fine-tuned for text generation on two language models. We will provide the experiment setup in the following as we proceed. Details on the training hyperparameters are deferred to Appendix \ref{sec:training}.

\subsubsection{Unlearning Benchmarks}
\label{unleanring_benchmarks} 

We adopted \sys and other MIAs to attack and evaluate the following SOTA unlearning methods.

\vspace{0.05in}

\noindent\textbf{$\ell_1$ Sparse} \cite{sparsity} based unlearning approach employs a \textit{pruning and fine-tuning} strategy. Initially, it removes weights with the smallest absolute values, based on the assumption that these weights contribute the least to the model's performance. This pruning step not only facilitates unlearning but also results in a sparser, potentially smaller neural network. After pruning, the model is fine-tuned on $D_r$ to recover its performance. 

\vspace{0.05in}
\noindent\textbf{Scrub}~\cite{scrub} employs a distillation-based approach for unlearning. It guides the model to diverge from the original predictions for $D_f$. 
Simultaneously, it ensures that the model's predictions on $D_r$ remain consistent with its original behavior. 
By leveraging distillation, Scrub achieves a balance between unlearning the forget data and preserving the accuracy and functionality of the remain data.

\vspace{0.05in}
\noindent\textbf{GA/GA+} \cite{thudi2022unrolling, scrub, golatkar2020eternal} is an enhanced version of gradient ascent tailored for unlearning. It operates in two stages: first, it maximizes the loss on $D_f$, then, it fine-tunes the model to minimize the loss on the $D_r$, ensuring that the model maintains its predictive accuracy on remain data.  We specifically add fine-tuning (refining) steps to make gradient ascent comparable with other benchmarks.

\vspace{0.05in}
\noindent\textbf{NegGrad+}~\cite{hayes2024inexact, scrub} is designed to optimize two goals: raising the loss on the forget set while reducing the loss on the remain data, controlled by a balancing parameter.

\vspace{0.02in}

There are growing baselines not included in this paper, as the methods presented above consistently achieved superior performance, in line with the findings of~\cite{hayes2024inexact, sparsity}. However, any inexact unlearning algorithm can be measured by \sys.

%% file: evaluation/evaluation_efficacy.tex
\subsubsection{Target Samples for Unlearning}
\label{mia_setting}
We collected target samples for unlearning by conducting classic Membership Inference Attacks (MIAs), specifically LiRA~\cite{carlini_lira}. 
For a given sample, $p(In)$ is the estimated likelihood probability that it was in the training set, while $p(Out)$ is the estimated probability that it was not. 
Samples with likelihoods exceeding a certain threshold at TPR@lowFPR are marked as highly memorized (i.e., vulnerable). 
Conversely, when $p(In)\approx p(Out)$, the samples are considered less memorized, making them more resistant to MIAs. Specifically, for CIFAR-10 and CIFAR-100, we consider the standard setting from~\cite{carlini_lira}, where a model is trained on a dataset containing half of the samples from training set (i.e., 25,000 training samples). 
The remain data serves as the attack dataset in Section \ref{mia_setting}. We then identified the vulnerable highly memorized samples (easier to attack) and protected samples as the ones with lower memorization (harder to attack). We refer readers to more details in \ref{details_vul_selection}

\vspace{0.05in}

\noindent\textbf{Target Samples for Shadow Model Training and Unlearning}. We select different setups of target samples and name these settings in the parentheses: 1) 600 random samples (``Random''), 2) 600 vulnerable samples (``Vulnerable''), 3) 600 protected samples (``Protected''), 4) 600 protected samples + 600 vulnerable samples (``Vulnerable + Protected''), and 5) 600 random samples of one class of dataset (``Class'').\footnote{Hayes et al.~\cite{hayes2024inexact} considers this setting as a challenging setting for CIFAR-10; We have excluded class-wise unlearning as removing an entire class from the dataset violates typical MI assumptions (blind attacks would outperform any targeted attack in such a setting).}

\subsubsection{MIA Settings}\label{mia_setting}
After performing model unlearning, we evaluate the unlearned models using MIAs.

\vspace{0.05in}
\noindent\textbf{Shadow Model Training.} To facilitate MIAs, we construct a training set that includes unlearned examples from the various unlearning settings, along with randomly sampled data from the attack dataset, which is non-overlapping with the unlearning set. 

For MIA, we trained 90 shadow models, each for 50 epochs.
The shadow models achieve an accuracy of at least 88\%. 
Additionally, we tune hyperparameters to ensure that the unlearning baselines match the \textit{Retrain} gold standard in accuracy across the remain, forget, and test datasets~\cite{retrain}. This involves conducting a comprehensive grid search similar to~\cite{hayes2024inexact}, but with a set of parameters specifically adapted to each target data setup (see~\autoref{tab:unleanring-baselines-param} in Appendix \ref{sec:unleanring-baselines-param} for details). Our implementation leverages the open-sourced code bases from these benchmarks.

\vspace{0.05in}
\noindent\textbf{Average-case Attack.} To implement an average-case attack as a baseline, we aggregate the logit-scaled confidences across all shadow models following similar population attack adaptations ~\cite{shokri2017membership, graves2021amnesiac, hayes2024inexact}. Specifically, for each setting (e.g., when the target data consists of vulnerable samples), we combine the outputs from all shadow models to form aggregated populations. A linear binary classifier is then fitted to these aggregated distributions, learning to differentiate between member and non-member populations across the entire set.

\subsection{Privacy: (Stand-alone) Unlearned Model}

\subsubsection{Privacy Leakage across Different Target Sets}
\label{sec:diff_target}

We evaluate \sys under the standard setting that we have shadow models for all target sample setups. This study helps illustrate which unlearning settings are more prone to our privacy attacks. As shown in~\autoref{tab:scrub_NegGrad}, we report Attack Accuracy, Attack AUC, and TPR@1\% FPR as privacy leakage metrics for different unlearning methods, considering different combinations of forget and target data. Additionally, we present the accuracy gap between the unlearned model ($\theta_\mathcal{U}$) and the retrained model ($\theta_\mathcal{R}$) to provide a general performance evaluation of the unlearning methods.

\vspace{-0.1in}

\input{Tables/scrub_attack}

The results show that privacy leakage is more pronounced when the forget set includes a mix of vulnerable and protected samples, compared to using random samples or random samples from ``Class''. For example, in GA+, when the forget data and target data consist of random samples or random samples from ``Class'', the attack AUC is 56\% and 61\%, respectively, while TPR@1\% FPR is 7\% and 3\%. Similarly, the attack accuracy is 54.75\% and 55.25\%. 
However, when the forget data consists of both Vulnerable and Protected samples and the target data is Vulnerable, the attack AUC, attack accuracy, and TPR@1\% FPR increase significantly to 76.17\%, 68.5\%, and 22.75\%, respectively. A similar trend is observed for Scrub and NegGrad+, indicating that \sys is highly effective at inducing privacy leakage across most unlearning methods. We attribute this behavior to the batch-averaging effect inherent in methods like NegGrad+, GA+, and Scrub, which optimize an objective loss on $D_f$ (distillation for Scrub and cross-entropy for NegGrad+ and GA+). As expected, $\ell_1$ Sparse demonstrates the highest resilience to \sys, likely due to its reduced memorization capacity, which makes it less vulnerable to both \sys and traditional MIAs \cite{sparse-mia-2} on model training. However, this resilience comes at a cost, as sparse models also exhibit reduced memorization of remaining vulnerable samples, a tradeoff discussed further in Section~\ref{experiment_efficacy}. 
Overall, we conclude that using a combination of Vulnerable + Protected samples as forget data and Vulnerable samples as target data represents the most effective setting for evaluating privacy leakage. 

Additionally, we observe that GA+ achieves a similar accuracy gap (7.75\% vs 7.5\%) compared to Scrub but exhibits a significantly higher TPR@1\% FPR, with a margin of 14.22\%. This highlights the inadequacy of relying on average accuracy to assess unlearning privacy leakage.

\vspace{-0.1in}

\begin{figure}[h]
    \centering
        \begin{subfigure}[t]{\linewidth}
        \centering
        \begin{subfigure}[t]{0.5\linewidth}
            \centering
            \includegraphics[width=\linewidth]{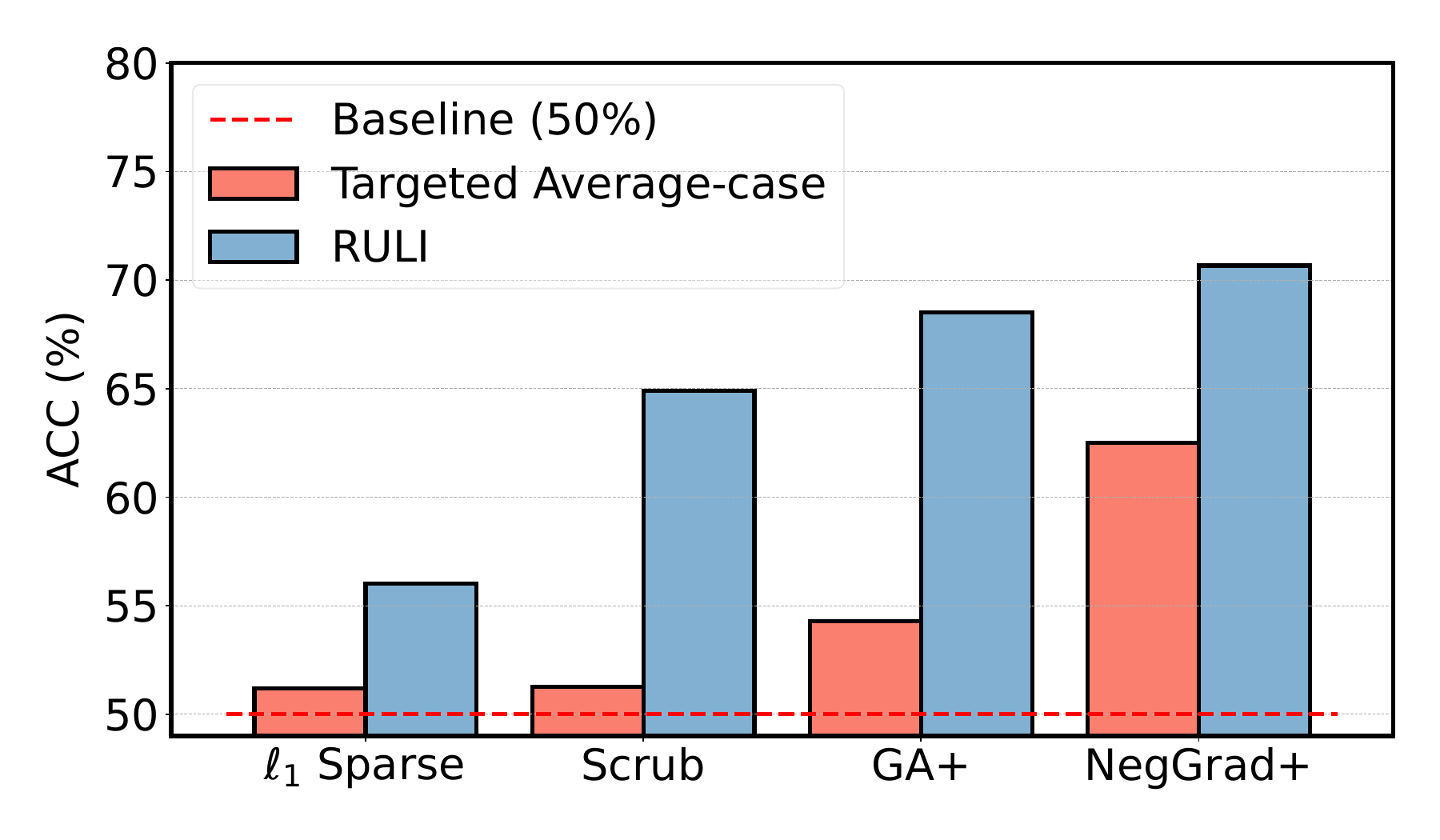}
        \end{subfigure}
        \hspace{-0.1in}
        \begin{subfigure}[t]{0.5\linewidth}
            \centering
            \includegraphics[width=\linewidth]{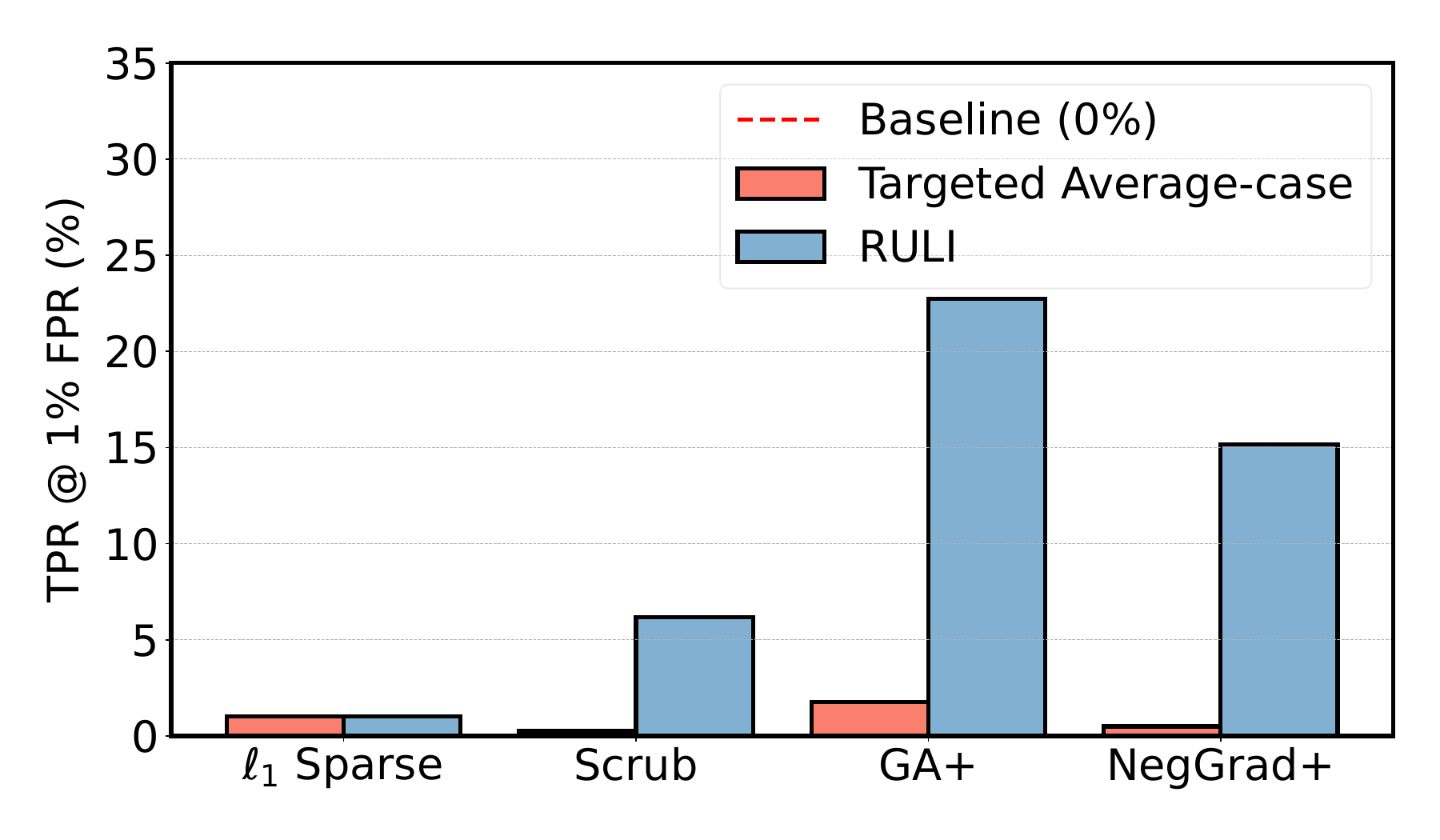}
        \end{subfigure}
        \caption{CIFAR-10 Results}
        \label{fig:cifar10_results}
    \end{subfigure}

   \vspace{0.15in} 


    \begin{subfigure}[t]{\linewidth}
        \centering
        \begin{subfigure}[t]{0.5\linewidth}
            \centering
            \includegraphics[width=\linewidth]{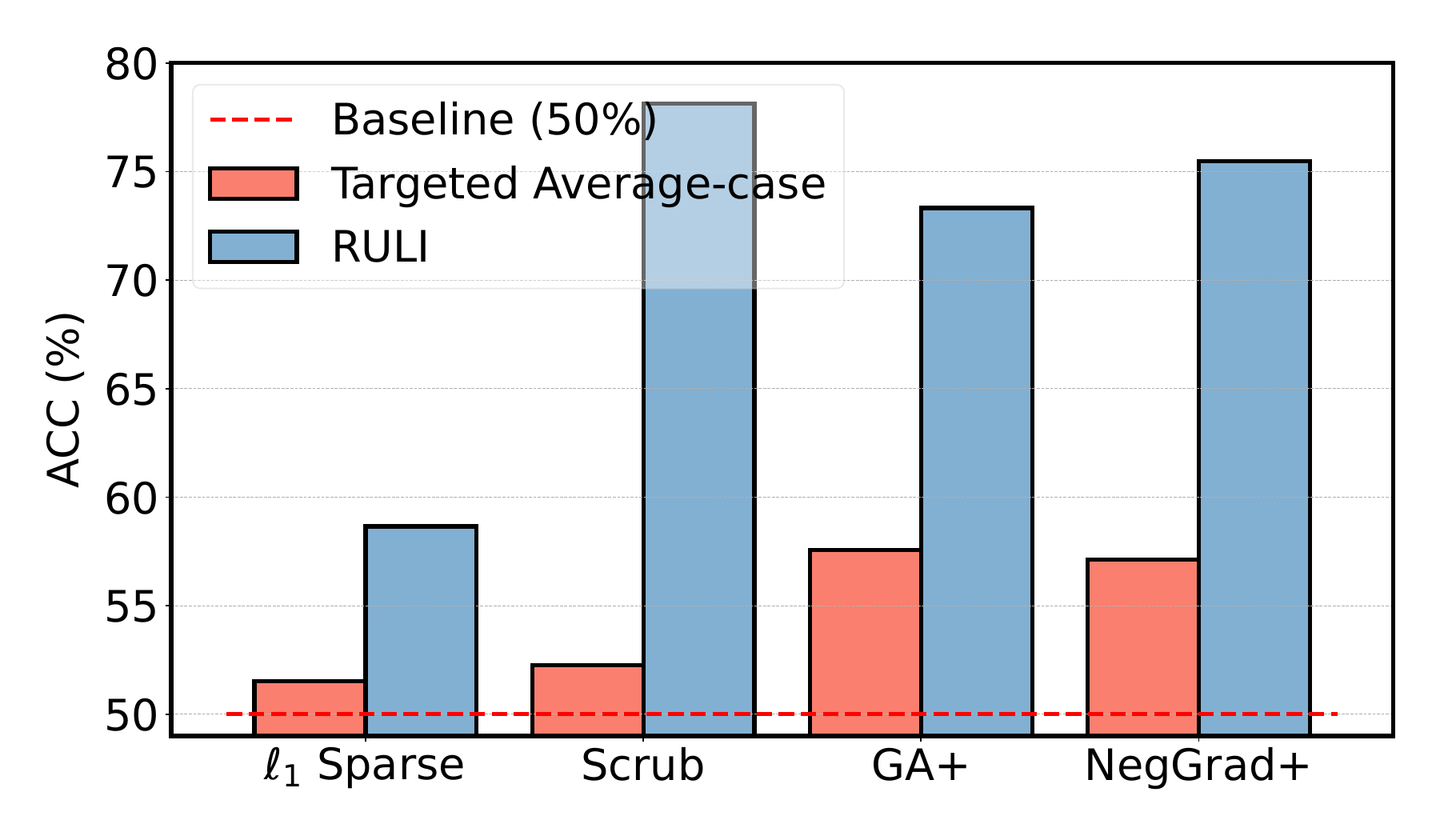}
        \end{subfigure}
        \hspace{-0.1in}
        \begin{subfigure}[t]{0.5\linewidth}
            \centering
            \includegraphics[width=\linewidth]{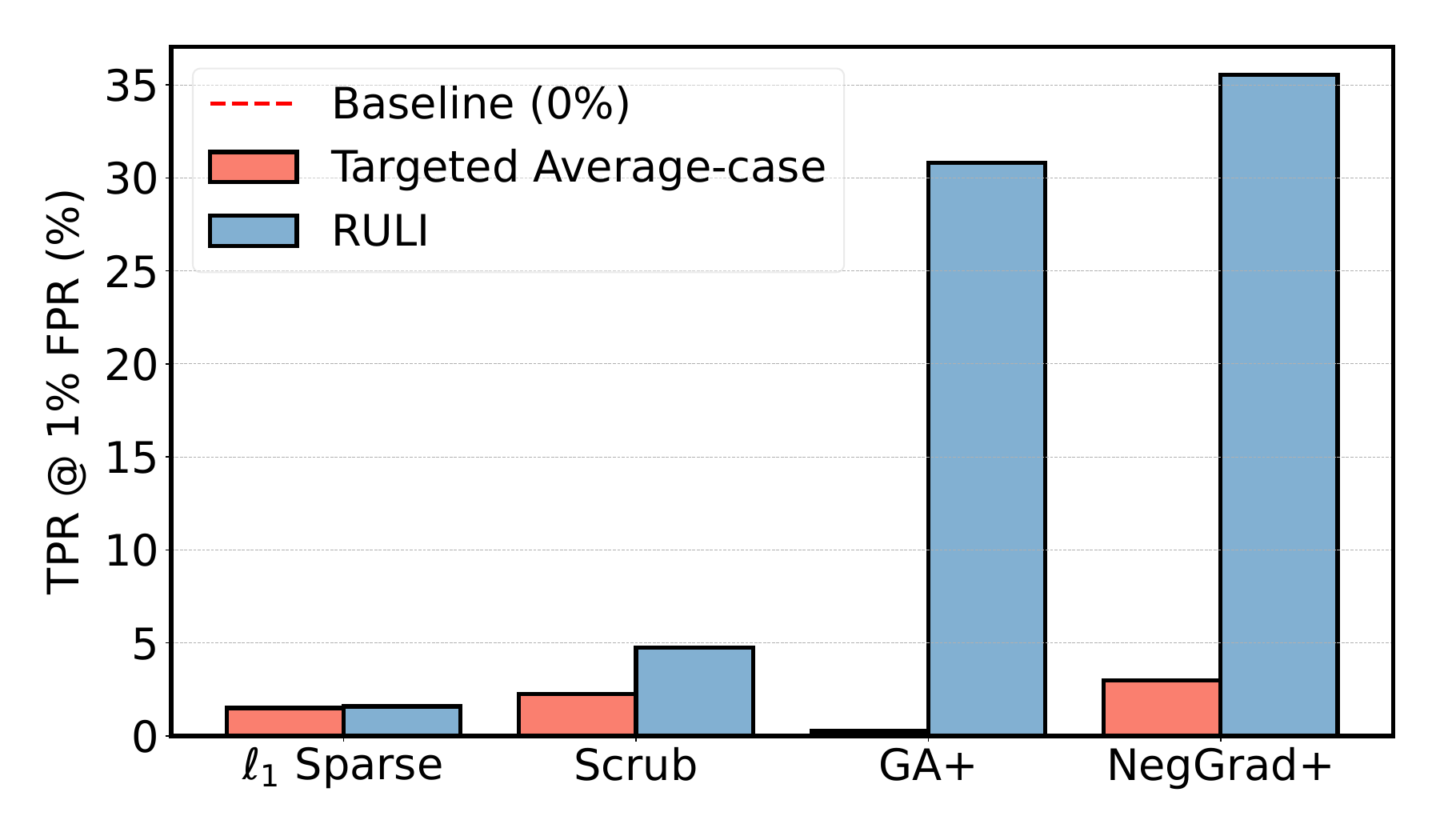}
        \end{subfigure}
        \caption{CIFAR-100 Results}
        \label{fig:cifar100_results}
    \end{subfigure}
    \vspace{-0.05in}
    \caption{Attack accuracy and TPR@1\%FPR targeting vulnerable samples as canary with \sys vs targeted average-case attack. Average-case attacks are underestimating the privacy risk (TPR@1\%FPR) even under target vulnerable samples. \sys achieves higher attack success on all methods. 
    }\vspace{-0.15in}
    \label{fig:auc_tpr_cifar10_100}
\end{figure}


\subsubsection{Vulnerable Samples as Canaries} 
\label{canary_setting}
We examine vulnerable samples analogous to canaries~\cite{one_line, aerni2024evaluations} in a similar setting. We inject 600 vulnerable samples (100 would be trained and remain in the model to prevent possible \textit{Onion effect}~\cite{onion}) along with 600 random samples, except for the vulnerable set. Consequently, we will have \textit{500 canaries to target (250 unlearned and 250 held-out)}.

\vspace{0.05in}
\noindent\textbf{Targeted Average-case Attack}. As a baseline vs \sys, we adapt average-case attacks to evaluate its performance. Specifically, we derive two populations from all distributions generated by shadow models for the vulnerable samples. Using these populations, we train a regression model to classify the samples based on their population membership. The model can distinguish between samples that were unlearned and those that were entirely held out from both the training and unlearning processes. This approach shows \sys's effectiveness in capturing nuanced privacy leakage. 
\vspace{0.05in}

\noindent \textbf{Results.} \autoref{fig:auc_tpr_cifar10_100} presents the attack accuracy and TPR@1\% FPR for CIFAR-10 and CIFAR-100 datasets, where \sys outperforms average-case attacks significantly. 
For CIFAR-10, across the unlearning benchmarks, the attack accuracy and TPR@1\% FPR achieved by \sys are, on average, 18.63\% and 11.88 times higher, respectively, compared to the average-case attacks. Similarly, for CIFAR-100, \sys surpasses average-case attacks by 30.72\% in attack accuracy and 9.38 times in TPR@1\% FPR. 
Thus, the average-case attacks significantly underestimate privacy risks, even when targeting vulnerable samples, underscoring the severity of \textbf{Pitfall I}.

We acknowledge that our strategy does not offer the tightest privacy leakage estimation. For example, leveraging only the top vulnerable samples and getting the membership inference over hundreds of runs on those samples, \cite{one_line, aerni2024evaluations} can potentially provide higher TPR@low FPR results. We leave this exploration to future works, positioning our findings as a step towards systematically attacking unlearning algorithms.

\begin{figure*}[!h]
    \centering
    \begin{subfigure}[b]{0.252\textwidth}
        \centering
        \includegraphics[width=\textwidth]{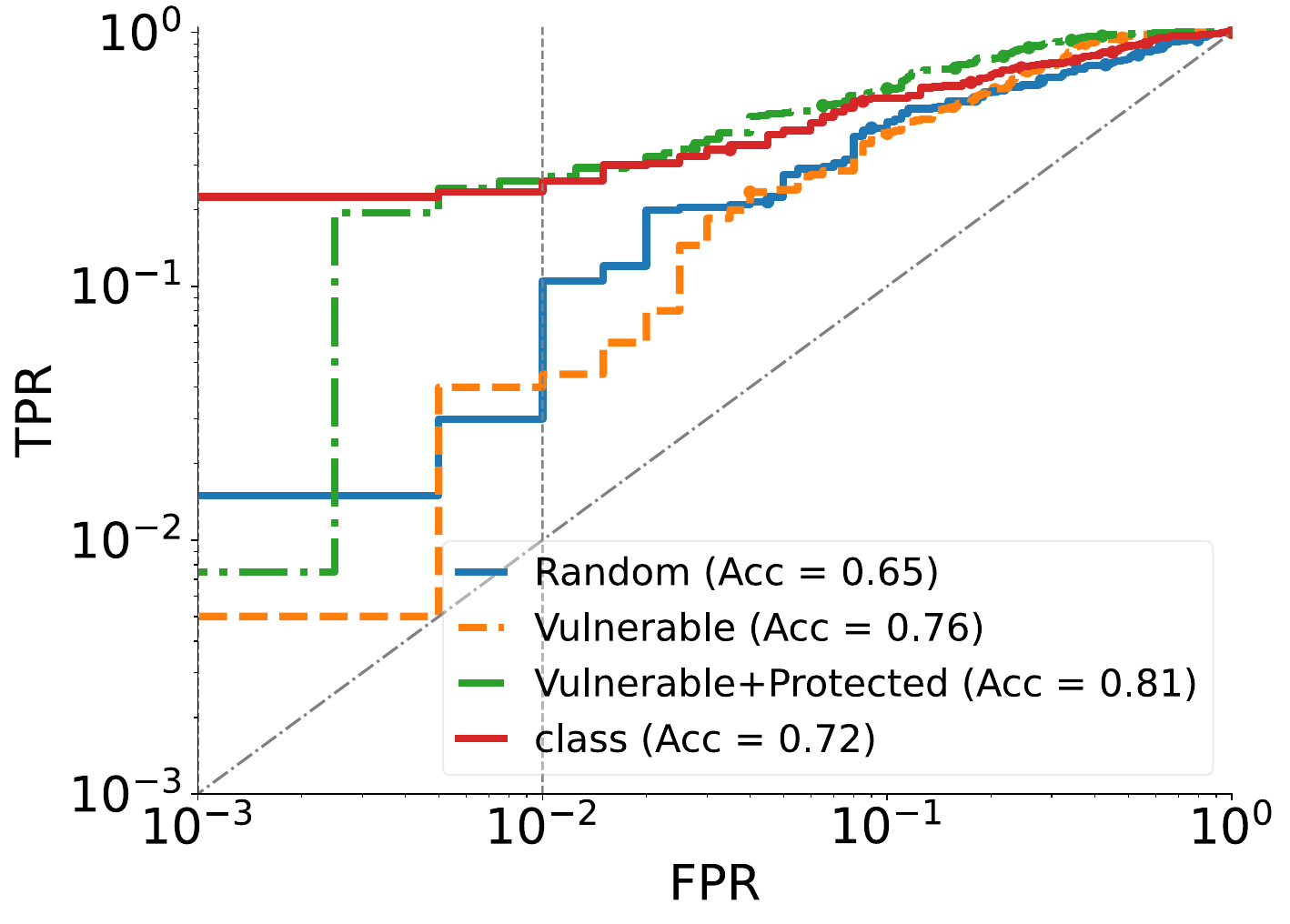}
        \caption{$\ell_1$ Sparse}
        \label{fig:sparse_roc}
    \end{subfigure}
    \hspace{-0.09in}
    \begin{subfigure}[b]{0.252\textwidth}
        \centering
        \includegraphics[width=\textwidth]{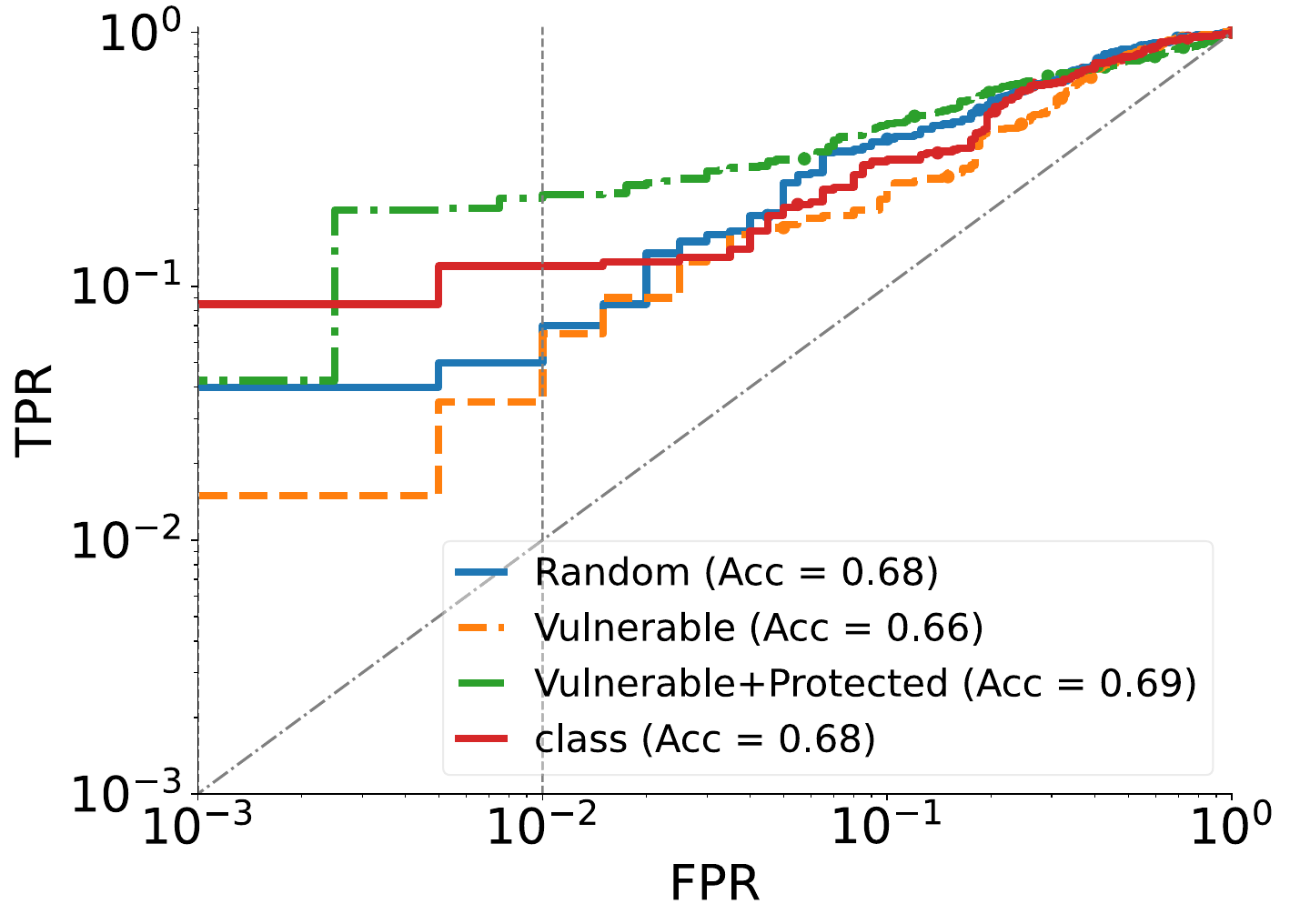}
        \caption{Scrub}
        \label{fig:scrub_roc}
    \end{subfigure}
    \hspace{-0.09in}
    \begin{subfigure}[b]{0.252\textwidth}
        \centering
        \includegraphics[width=\textwidth]{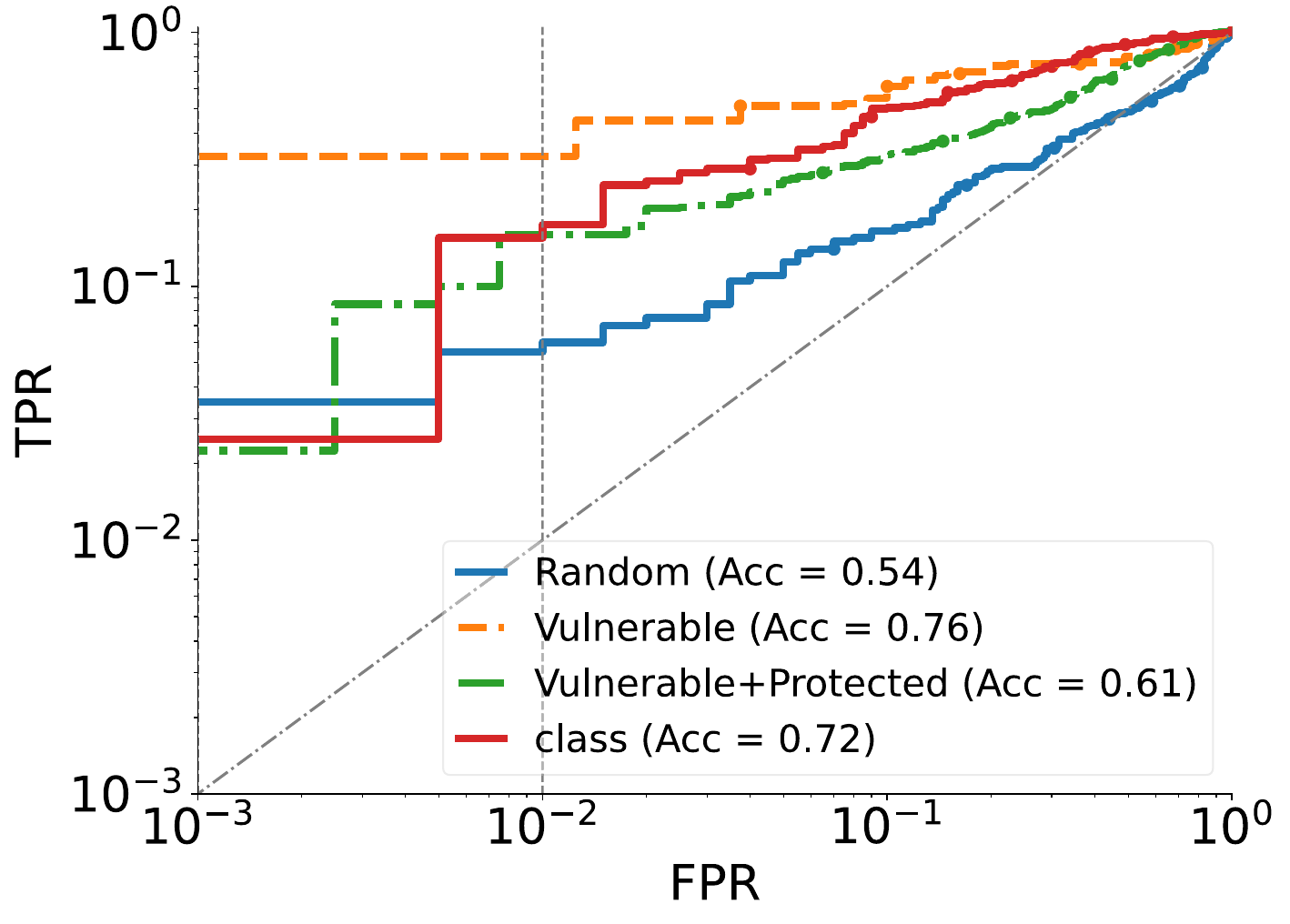}
        \caption{GA+}
        \label{fig:ga_roc}
    \end{subfigure}
    \hspace{-0.09in} 
    \begin{subfigure}[b]{0.252\textwidth}
        \centering
        \includegraphics[width=\textwidth]{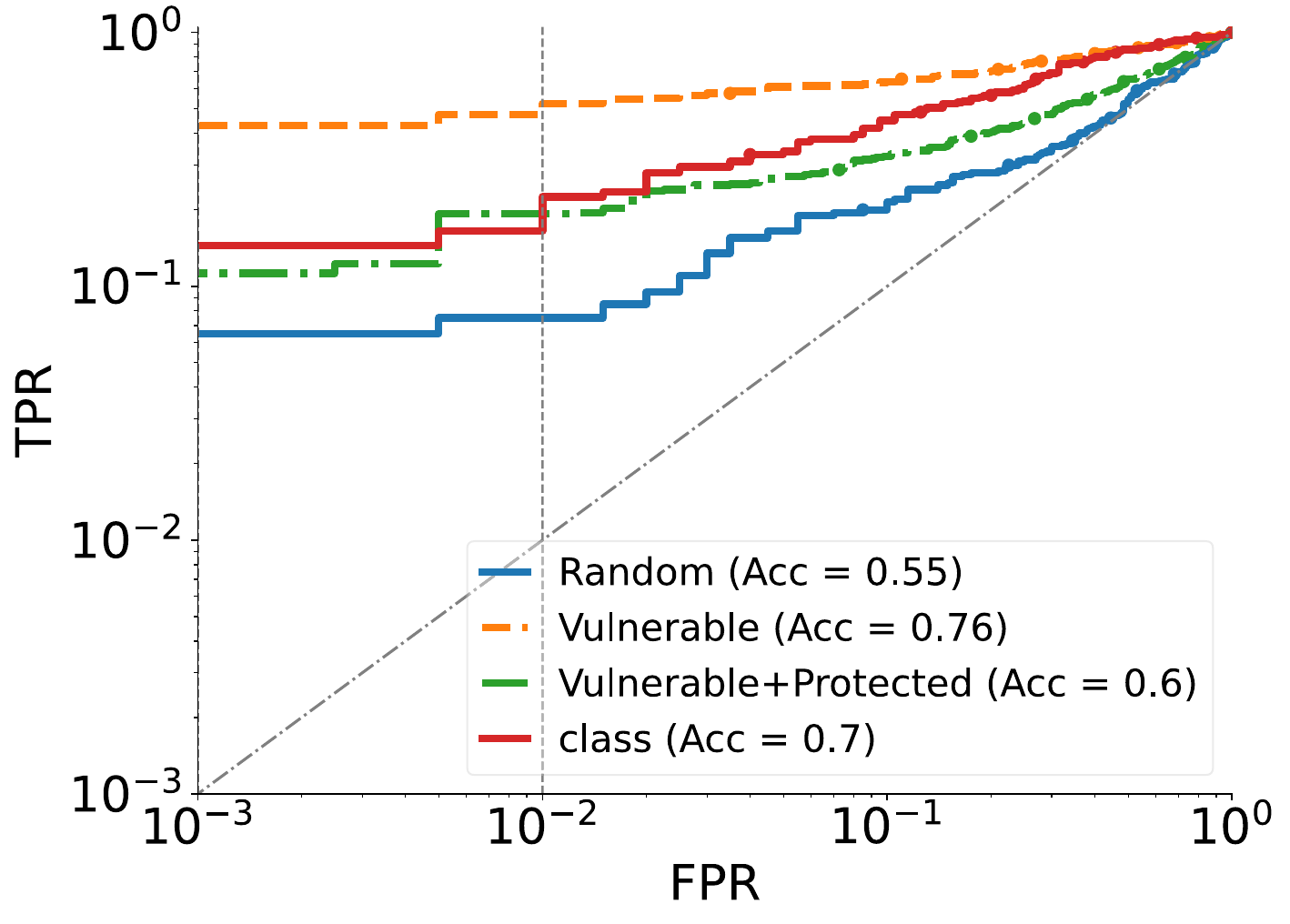}
        \caption{NegGrad+}
        \label{fig:negrad_roc}
    \end{subfigure}
    \vspace{-0.2in}
    \caption{Comparing with retraining using \sys. All unlearning benchmarks are highly distinguishable from retraining.}\vspace{-0.1in}
    \label{fig:retrained_unlearned}
\end{figure*}

\subsection{Efficacy: Unlearning vs. Retraining}
\label{experiment_efficacy}

We evaluated \sys for efficacy attack to determine whether a target sample was unlearned or retrained, as described in Game~\ref{game3}. 

To achieve this, we utilized Algorithm \ref{alg:lira} to compute the likelihood test ratio in accordance with ~\autoref{eq:test_efficacy}. 
This experiment requires the Test model to construct a hypothetical test model, as defined in~\autoref{eq:test_function} and illustrated in~\autoref{fig:revision_framework}. 
The goal of this experiment is to provide a more fine-grained evaluation for comparing the effectiveness of inexact unlearning methods against the retraining global standard.

We consider four attack settings, where the target data consist of ``Random'' Samples, ``Vulnerable'' Samples, ``Vulnerable + Protected'' Samples, and randomly selected samples from ``Class'', respectively.~\autoref{fig:sparse_roc}-\ref{fig:negrad_roc}, demonstrating that by querying the test model $\mathcal{\theta_T}$, the attacker achieves a high attack success rate, effectively distinguishing many samples under TPR@FPR.
From the figures, we observe that in our newly introduced attack settings (Vulnerable and Vulnerable + Protected), TPR@FPR is larger (TPR@1\%FPR$\geq$10\%) across all unlearning methods. 
Notably, the Vulnerable + Protected setting dominates on the $\ell_1$ Sparse and Scrub. The Vulnerable setting achieves better results on GA+ and NegGrad+. These findings indicate that \sys successfully avoids \textbf{Pitfall III}, effectively distinguishing retrained models from unlearned models. While these results do not show privacy leakage, they establish a promising foundation also for future works. One could leverage this method to audit the upper bounds of certified unlearning approaches ~\cite{guo2019certified, certified2, cetified1}.

\vspace{0.05in}
\noindent\textbf{Some Insights.} Efficacy attack can be used as the definitive evaluation standard, as it provides a clear basis for comparing inexact unlearning with retraining. An interesting finding that is not observable in average-case accuracy evaluation is the unintended loss of memorization of remained vulnerable samples caused by inexact unlearning algorithms. Ideally, only the information related to the forget set should be removed. However, we observed that the model's accuracy on the remaining vulnerable dataset was also unintentionally reduced.

\vspace{-0.1in}

\begin{table}[!ht]
\small
\centering
\caption{Inexact unlearning methods cannot accurately preserve the remained vulnerable samples (shown as Vul set).}
\vspace{-0.1in}
\resizebox{\linewidth}{!}{%
\begin{tabular}{@{}l|cccc|cccc@{}}
\toprule
\multirow{2}{*}{\begin{tabular}[c]{@{}l@{}} \\Unlearning \\ Methods\end{tabular}} & \multicolumn{4}{c|}{CIFAR-10} & \multicolumn{4}{c}{CIFAR-100} \\ \cmidrule(l){2-9} 
 & $Acc(D_f$) & $Acc(D_r)$ & $Acc(D_\text{test})$ & \begin{tabular}[c]{@{}c@{}}$Acc(D_r)$\\ on Vul set\end{tabular} & $Acc(D_f$) & $Acc(D_r)$ & $Acc(D_\text{test})$ & \begin{tabular}[c]{@{}c@{}}$Acc(D_r)$\\ on Vul set\end{tabular} \\ \midrule
$\ell_1$ Sparse & 57.5\% & 94.84\% & 88.36\% & \textbf{45.10\%} & 51\% & 87.48\% & 66.34\% & \textbf{50.54\%} \\
Scrub & 65.0\% & 97.58\% & 89.63\% & 70.10\% & 61.5\% & 96\% & 64.86\% & 83.70\% \\
GA+ & 66.60\% & 94.42\% & 84.24\% & 62.50\% & 67\% & 99.11\% & 63.13\% & 100\% \\ 
NegGrad+ & 81.0\% & 96.93\% & 85.74\% & 80.96\% & 65.0\% & 91.89\% & 58.41\% & 86.96\% \\ \midrule
Retrain & 59.50\% & 99.79\% & 90.71\% & 96.19\% & 35.75\% & 99.93\% & 67.39\% & 99\% \\ \bottomrule
\end{tabular}
}\vspace{-0.05in}
\label{table:inexact_unlearning}
\end{table}

As shown in~\autoref{table:inexact_unlearning}, $ACC(D_f)$, $ACC(D_r)$, $ACC(D_\text{test})$, and $ACC(D_r)$ on Val set denote accuracy on forget data, remain data, test data, and remaining vulnerable data, respectively. Generally, after unlearning, the model performs well on forget data, remain data, and test data, as the $ACC(D_f)$, $ACC(D_r)$, $ACC(D_\text{test})$ of unlearned models are comparable to those of retrained models for both datasets. However, unlearned models show a significant accuracy drop on remaining vulnerable data compared to the retrained model. For CIFAR-10, $ACC(D_r)$ on the vulnerable set is 96.19\% for the retrained model but only 64.67\% on average for unlearned models. Similarly, for CIFAR-100, $ACC(D_r)$ on Val set is 99\% for the retrained model versus 80.3\% on average for unlearned models.

Among the benchmarks, $\ell_1$ Sparse exhibited the most pronounced loss of memorization on the remain data. We hypothesize that this is due to the strong global sparsity enforced by the method. 
While the sparsity was maintained at approximately 60\% to optimize the accuracy gap, it likely contributed to the observed loss in memorization. Notice that, to validate our findings, we reduce model sparsity to 7\%, observing only a 1–2\% change in accuracy on the remaining vulnerable set.

%% file: Tables/scrub_attack.tex
\begin{table}[!h]
\caption{Attacking or evaluating unlearning methods on the CIFAR-10 dataset (whether by selecting safe samples, class-specific random samples, or exclusively vulnerable samples as forget sets, does not effectively assess privacy risk--\textbf{Pitfall II}). A better approach is to use target samples as canaries. Inferences are performed in a single training and unlearning run, selecting the best unlearning instance for evaluation.}
\vspace{-0.1in}
\centering
\resizebox{\linewidth}{!}{%
\begin{tabular}{@{}l|l|l|cccc@{}}
\toprule
\begin{tabular}[c]{@{}l@{}}Unlearning\\Methods\end{tabular} & Forget Data & Target Data & \begin{tabular}[c]{@{}c@{}}Attack\\ AUC\end{tabular} & \begin{tabular}[c]{@{}c@{}}Attack\\ ACC\end{tabular} & \begin{tabular}[c]{@{}c@{}}TPR@\\ 1\% FPR\end{tabular} & \begin{tabular}[c]{@{}c@{}}$\Delta$\\ ACC$(D_f)$\end{tabular} \\ \midrule
\multirow{5}{*}{$\ell_1$ Sparse} & + Random & Random & 45\% & 48\% & 1\% & 2.5\% \\ \cmidrule(l){2-7} 
 & + Class & Class & 54.0\% & \textbf{53.75\%} & 0.0\% & 7.5\% \\ \cmidrule(l){2-7} 
 & + Vulnerable & Vulnerable & 55.97\% & 50.50\% & 1\% & 3\% \\ \cmidrule(l){2-7} 
 & + (Vulnerable + Protected) & Protected & 48.96\% & 48.46\% & 1.06\% & \multirow{2}{*}{4\%} \\
 & + (Vulnerable + Protected) & Vulnerable & \textbf{56.85\%} & 53.37\% & \textbf{1.9\%} &  \\ \midrule
\multirow{5}{*}{Scrub} & + Random & Random & 52\% & 51.75\% & 3\% & 0.5\% \\ \cmidrule(l){2-7} 
 & + Class & Class & 61\% & 57\% & 1\% & 0.3\% \\ \cmidrule(l){2-7} 
 & + Vulnerable & Vulnerable & 57.0\% & 54.5\% & 5\% & 3.0\% \\ \cmidrule(l){2-7} 
 & + (Vulnerable + Protected) & Protected & 57.75\% & 49.98\% & 0.0\% & \multirow{2}{*}{7.5\%} \\
 & + (Vulnerable + Protected) & Vulnerable & \textbf{73.53\%} & \textbf{68.27\%} & \textbf{8.53\%} &  \\ \midrule
\multirow{5}{*}{GA+} & + Random & Random & 56\% & 54.75\% & 7\% & 10.5\% \\ \cmidrule(l){2-7} 
 & + Class & Class & 61\% & 55.25\% & 3\% & 7\% \\ \cmidrule(l){2-7} 
 & + Vulnerable & Vulnerable & 60.75\% & 65\% & 6\% & 10\% \\ \cmidrule(l){2-7} 
 & + (Vulnerable + Protected) & Protected & 52.13\% & 54.17\% & 0.0\% & \multirow{2}{*}{7.75\%} \\
 & + (Vulnerable + Protected) & Vulnerable & \textbf{76.17\%} & \textbf{68.5\%} & \textbf{22.75\%} &  \\ \midrule
\multirow{5}{*}{NegGrad+} & + Random & Random & 52\% & 52.75\% & 4\% & 2\% \\ \cmidrule(l){2-7} 
 & + Class & Class & 69\% & 58.50\% & 4\% & 17\% \\ \cmidrule(l){2-7} 
 & + Vulnerable & Vulnerable & 76\% & 70.25\% & 21\% & 43.5\% \\ \cmidrule(l){2-7} 
 & + (Vulnerable + Protected) & Protected & 52.5\% & 51.56\% & 1.59\% & \multirow{2}{*}{15.5\%} \\
 & + (Vulnerable + Protected) & Vulnerable & \textbf{77.54\%} & \textbf{72.36\%} & \textbf{18.48\%} &  \\ \bottomrule

\end{tabular}
}\vspace{-0.1in}
\label{tab:scrub_NegGrad}
\end{table}

%% file: Tables/revision_comparison.tex
\vspace{-0.05in}

\begin{table}[!h]
\caption{
Ablation study comparing U-LiRA and \sys under different target selection strategies and MIAs for Scrub and $\ell_1$ Sparse on CIFAR-10 dataset. \textit{For ``Vulnerable + Protected'', privacy leakage is computed only on vulnerable samples.} Results are averaged over 10 independent runs.} 
\vspace{-0.1in}
\label{tab:revision_compare_ulira}
\resizebox{\linewidth}{!}{%
\begin{tabular}{@{}lcccccc@{}}
\toprule
\multicolumn{1}{c}{} &  &  & \multicolumn{2}{c}{\textbf{Privacy Leakage}} & \multicolumn{2}{c}{\textbf{MIA for Efficacy}} \\ \cmidrule(l){4-7} 
\multicolumn{1}{c}{\multirow{-2}{*}{\textbf{Setting}}} & \multirow{-2}{*}{\textbf{Target Samples}} & \multirow{-2}{*}{\textbf{MIA Method}} & ACC $\uparrow$ & \multicolumn{1}{c|}{\begin{tabular}[c]{@{}c@{}}TPR@1\\ \%FPR $\uparrow$\end{tabular}} & ACC $\uparrow$ & \begin{tabular}[c]{@{}c@{}}TPR@1\\ \%FPR $\uparrow$\end{tabular} \\ \midrule

\multicolumn{7}{l}{\cellcolor[HTML]{C0C0C0}{\color[HTML]{000000} \textit{$\ell_1$ Sparse}}} \\
U-LiRA Baseline & one "Class" only & \multicolumn{1}{c|}{U-LIRA} & 50.2\% & \multicolumn{1}{c|}{0.5\%} & - & - \\
Alt. MIA Only & Vulnerable + Protected & \multicolumn{1}{c|}{U-LiRA} & 55.7\% & \multicolumn{1}{c|}{1.6\%} & - & - \\
Alt. Target Only & one "Class" only & \multicolumn{1}{c|}{\sys} & 51.73\% & \multicolumn{1}{c|}{1.37\%} & 65.6\% & 18.0\% \\
\sys (Full) & Vulnerable + Protected & \multicolumn{1}{c|}{\sys} & 54.4\% & \multicolumn{1}{c|}{1.4\%} & 81.1\% & 23.3\% \\

\midrule

\multicolumn{7}{l}{\cellcolor[HTML]{C0C0C0}{\color[HTML]{000000} \textit{Scrub}}} \\
U-LiRA Baseline & one "Class" only & \multicolumn{1}{c|}{U-LIRA} & 53.75\% & \multicolumn{1}{c|}{1.13\%} & - & - \\
Alt. MIA Only & Vulnerable + Protected & \multicolumn{1}{c|}{U-LiRA} & 62.8\% & \multicolumn{1}{c|}{8.6\%} & - & - \\
Alt. Target Only & one "Class" only & \multicolumn{1}{c|}{\sys} & 53.32\% & \multicolumn{1}{c|}{1.7\%} & 65.1\% & 7.0\% \\
\sys (Full) & Vulnerable + Protected & \multicolumn{1}{c|}{\sys} & 65.5\% & \multicolumn{1}{c|}{11.8\%} & 67.4\% & 8.8\% \\

\bottomrule
\end{tabular}
}\vspace{-0.05in}
\end{table}

%% file: Tables/vit_experiments.tex
\begin{table}[!h]
\centering
\caption{Privacy leakage on TinyImageNet. We evaluate three target selection strategies—\textit{Canary injection}, \textit{Vulnerable-only}, and \textit{Random samples}—using a target set of 500 samples (250 unlearned and 250 are held-out). All experiments are performed on a Swin-small ViT model.}
\label{table:vit_experiments}
\resizebox{\linewidth}{!}{%
\begin{tabular}{@{}ccccccccc@{}}
\toprule
 & \multicolumn{4}{c|}{\begin{tabular}[c]{@{}c@{}}Targeted average-case attack \\ (Population attack)\end{tabular}} & \multicolumn{4}{c}{RULI} \\ \cmidrule(l){2-9} 
\multirow{-2}{*}{Target data} & AUC & ACC & \begin{tabular}[c]{@{}c@{}}TPR@\\ 1\%FPR\end{tabular} & \multicolumn{1}{c|}{\begin{tabular}[c]{@{}c@{}}TPR@\\ 5\%FPR\end{tabular}} & AUC & ACC & \begin{tabular}[c]{@{}c@{}}TPR@\\ 1\% FPR\end{tabular} & \begin{tabular}[c]{@{}c@{}}TPR@\\ 5\%FPR\end{tabular} \\

\midrule

\multicolumn{9}{l}{\cellcolor[HTML]{C0C0C0}{\color[HTML]{000000} \textit{$\ell_1$ Sparse}}} \\ 
\begin{tabular}[c]{@{}c@{}}Vulnerable\\ only\end{tabular} & 54.4\% & 55.1\% & 2.3\% & \multicolumn{1}{c|}{5.2\%} & 59.6\% & 56.0\% & 2.4\% & 12.4\% \\
\begin{tabular}[c]{@{}c@{}}Vulnerable\\ as canaries\end{tabular} & 55.3\% & 54.7\% & 0.8\% & \multicolumn{1}{c|}{5.6\%} & \textbf{62.6\%} & \textbf{57.0\%} & \textbf{6.3\%} & \textbf{16.6\%} \\
Random & 53.2\% & 52.8\% & 0.0\% & \multicolumn{1}{c|}{2.4\%} & 56\% & 54.4\% & 0.8\% & 6.4\% \\

\midrule

\multicolumn{9}{l}{\cellcolor[HTML]{C0C0C0}\textit{Scrub}} \\ 
\begin{tabular}[c]{@{}c@{}}Vulnerable\\ only\end{tabular} & 52.5\% & 52.4\% & 2.0\% & \multicolumn{1}{c|}{5.4\%} & 65.3\% & 61.5\% & 11.7\% & 23.9\% \\
\begin{tabular}[c]{@{}c@{}}Vulnerable\\ as canaries\end{tabular} & 56.0\% & 56.2\% & 1.0\% & \multicolumn{1}{c|}{6.3\%} & \textbf{69.5\%} & \textbf{63.6\%} & \textbf{10.9\%} & \textbf{27.1\%} \\
Random & 49.6\% & 49.8\% & 1.0\% & \multicolumn{1}{c|}{2.8\%} & 59.7\% & 57.0\% & 6.0\% & 14.0\% \\

\bottomrule
\end{tabular}
}\vspace{-0.05in}
\end{table}

%% file: Discussion.tex
\section{Related Works}  
\vspace{-0.05in}
\noindent\textbf{Machine Unlearning}. Numerous studies have explored machine unlearning, broadly categorized into data-oriented and model-oriented approaches~\cite{10.1145/3603620}.

First, data-oriented methods focus on modifying training data. The SISA framework by Bourtoule et al.~\cite{bourtoule2021machine} partitions data into subsets, enabling selective retraining. Tarun et al.~\cite{tarun2023fast} proposed adding error-maximizing noise to obscure forgotten samples. Second, model-oriented methods modify the model directly. Golatkar et al.~\cite{golatkar2020eternal, baumhauer2022machine} used Fisher information~\cite{martens2020new} for selective forgetting, and later developed quadratic unlearning methods~\cite{golatkar2021mixed}, though at some cost to accuracy. Chundawat et al.~\cite{chundawat2023can} introduced a student-teacher approach, and their zero-shot Gated Knowledge Transfer (GRT) method~\cite{chundawat2023zero} uses pseudo data with a band-pass filter to isolate retained knowledge. Similar post-hoc model adjustment techniques have been developed for generative models. All inexact unlearning baselines in the evaluation fall in this category.

Furthermore, a parallel line of work investigates certified unlearning, which provides formal guarantees that unlearning approximates retraining without the removed data~\cite{guo2019certified, certified2}. These methods, inspired by differential privacy, define $(\epsilon, \delta)$-bounded differences in model outputs, often focusing on loss or gradient divergence. While theoretically sound, certified unlearning relies on conservative assumptions and tight constraints, making it less scalable to large models. Thus, we focus on evaluating practical, inexact unlearning methods, where empirical privacy risks are more pronounced.

\vspace{0.02in}

\noindent\textbf{Other Threats of Machine Unlearning}. Although machine unlearning is designed to enhance data privacy, recent work has uncovered significant privacy risks. Most existing attacks operate under a strict threat model. Chen et al.~\cite{chen2021machine, petsunlearning} showed that differences between models before and after unlearning can unintentionally leak information, enabling average-case MIA-based attacks to succeed. Under the same assumptions,~\cite{inversion_S&P2024} demonstrated model inversion attacks to further expose unlearning vulnerabilities. Attribute inference attacks~\cite{ganju2018property,ArevaloNDHW24}, which infer sensitive features from observed data, have also been adapted to unlearning. Stock et al.~\cite{stock2023lessons} investigated such attacks in the context of feature-level unlearning in white-box settings, evaluating whether specific attributes were present in the original dataset. Model inversion attacks~\cite{fredrikson2015model} pose additional risks by reconstructing input data from model outputs. These attacks are particularly effective against models subjected to class-level unlearning~\cite{baumhauer2022machine}. Graves et al.~\cite{graves2021amnesiac} advanced this line of work by adapting model inversion techniques to assess the robustness of class-level unlearning strategies. Other lines of attack involve injecting poisoned data or issuing adversarial unlearning requests to disrupt the unlearning process~\cite{huang2024unlearn}. These attacks aim to corrupt the unlearning mechanism and represent valuable, emerging directions; in contrast, our work focuses on evaluating whether unlearning has occurred under an assumed correct process.

%% file: conclusion.tex
\section{Conclusion}\vspace{-0.05in}
In this work, we have proposed \sys, a novel framework that fills critical gaps in evaluating privacy and efficacy of inexact unlearning. \sys employs a dual-objective attack to provide fine-grained insights, outperforming existing methods in distinguishing unlearned models from retrained models and exposing privacy risks in SOTA benchmarks. We have validated its effectiveness on image (e.g., CIFAR-10/100, TinyImageNet) and text (e.g., WikiText-103) datasets using ViT and language models. Our findings reveal persistent vulnerabilities, particularly for sensitive samples, and highlight the limitations of existing unlearnings. \sys sets a new evaluation standard for robust, scalable unlearning.

\section*{Acknowledgments}
 We sincerely thank the anonymous shepherd and all the reviewers for their constructive comments. This work is partially supported by the National Science Foundation (NSF) under Grants No. CNS-2432533, CNS-2308730, CNS-2302689, CNS-2319277, CNS-2241713, CNS-2331302, CNS-2339686, CNS-2029038, and CNS-2135988, as well as by a Cisco Research Award and the Synchrony Fellowship. Nima was also partially supported by the NSERC Discovery Program (Grant No. N01035) during his preliminary research on unlearning privacy in Canada, prior to beginning his Ph.D. at UConn.

%% file: Appendix.tex
\section*{Appendix} 

\begin{table}[ht]
\footnotesize\caption{Frequently used notations.}\vspace{-0.1in}
\resizebox{\linewidth}{!}{%
\begin{tabular}{@{}ll@{}}
\toprule
Symbol & Meaning \\ \midrule
$\theta_\mathcal{I}$ & Trained model \\
$\theta^*$ & Intermediate model parameter after t steps unlearning optimizations \\
$\theta_\mathcal{R}$ & Retrained model \\
$\theta_\mathcal{U}$ & Unlearned model \\
$\theta_\mathcal{T}$ &
Test model \\
${D_\text{train}}$ & Training data \\
$\mathcal{D}$ & Dataset Distribution \\
${D}_r$ & Remain data \\
${D}_f$ & Forget data \\
${D}_\text{target}$ & Target data \\
$z$ & Target sample \\
$\mathcal{Q}_h (z)$ & \begin{tabular}[c]{@{}l@{}}held-out distribution (from a sample was not in training \\ and unlearning\end{tabular} \\
$\mathcal{Q}_r (z)$ & \begin{tabular}[c]{@{}l@{}}remained distribution (from a sample was in training \\ but not in unlearning)\end{tabular} \\
$\mathcal{Q}_u (z)$ & \begin{tabular}[c]{@{}l@{}}unlearned distribution (from a sample was in training \\ and in unlearning)\end{tabular} \\
$\mathcal{Q}_i (z)$ & in distribution (from a sample was in training) \\
$\mathcal{Q}_R (z)$ & retrain distribution (from a sample retrained/not in training) \\
$\mathcal{A}$ & Training algorithm \\
$\mathcal{U}$ & Unlearning algorithm \\
$A$ & Adversary (attacker) \\
$C$ & Challenger \\
$N$ & Number of shadow models \\
\bottomrule
\end{tabular}
}\label{table:notation}%
\end{table}

\section{Further Implementation Details} 

\label{further_implementation_detailss}
\subsection{Training Details} \label{sec:training}

\begin{table}[h]
\small
\centering
\caption{Training configurations for different models/datasets}\vspace{-0.1in}
\label{table:training_config}
\resizebox{\linewidth}{!}{%
\begin{tabular}{@{}cccc@{}}
\toprule
\multirow{2}{*}{\textbf{Training Parameters}} & \textbf{CIFAR-10} & \textbf{CIFAR-100} & \textbf{TinyImageNet} \\ \cmidrule(l){2-4}
                                             & \textbf{ResNet-18} & \textbf{ResNet-18} & \textbf{Swin-small}  \\ \midrule
Training\slash fine-tuning epochs            & 50                 & 50                 & 2  \\
Batch size                                   & 128                & 128                & 128 \\
Momentum                                     & 0.9                & 0.9                & 0.9 \\
Learning rate                                & $1 \times 10^{-1}$ & $1 \times 10^{-1}$ & $1 \times 10^{-4}$ \\
Optimizer                                    & SGD                & SGD                & AdamW \\
$\ell_{2}$ regularization                    & $5 \times 10^{-4}$ & $5 \times 10^{-4}$ & $5 \times 10^{-4}$ \\ \bottomrule
\end{tabular}
}%
\end{table}

\begin{table}[ht]
\small
\centering
\caption{Training configurations for language model unlearning using SFT with prefix language modeling.}
\vspace{-0.1in}
\label{table:training_config-LM}
\begin{tabular}{@{}lc@{}}
\toprule
\textbf{Training Parameters} & \textbf{Configuration} \\ \midrule
Supervised training epochs              & 5 \\
Batch size                   & 16 \\
Learning rate                & $5 \times 10^{-5}$ \\
Optimizer                    & AdamW \\
Prefix language modeling     & Enabled (+2 epochs) \\
Early Stopping               & Enabled \\
Weight decay                 & 0.01 \\
\bottomrule
\end{tabular}
\end{table}

The training configurations for different models and datasets involved in the experiments are shown in~\autoref{table:training_config}. All models were developed using PyTorch \cite{NEURIPS2019_bdbca288} for image classification; for text generation, we used Hugging Face \footnote{
\href{https://huggingface.co/collections/EleutherAI/pythia-scaling-suite-64fb5dfa8c21ebb3db7ad2e1}{Pythia on Hugging Face}, \href{https://huggingface.co/openai-community/gpt2}{GPT-2 on Hugging Face}
} implementations of the language models for training and unlearning. 
The experiments are performed on servers equipped with multiple NVIDIA H100 Hopper GPUs (80 GB each).

\subsection{Hyperparameters of Unlearning Baselines} \label{sec:unleanring-baselines-param}

\input{Tables/baseline_param}

We carefully tuned the hyperparameters to ensure that the unlearning baselines are close to the accuracy performance of the Retrain gold standard. We assumed the model provider can adapt the parameters adaptively according to the forget data (though in practice, this might be challenging). Specifically, we conducted a comprehensive grid search, tuning the parameter sets to each unlearning setting.~\autoref{tab:unleanring-baselines-param} presents the hyperparameters used for various unlearning settings.


\subsection{Target Data}
\label{details_vul_selection}
\textbf{Details on Identifying Vulnerable and Protected Samples}. We train 256 shadow models on the dataset and set TPR@0.01\% FPR as the threshold for classifying highly memorized samples. 
In this scenario, the attacker's confidence is quantified by the likelihood ratio ($\tau = \frac{p(\emph{In})}{p(\emph{In})+p(\emph{Out})}$), which, if it hovers around $0.5\pm 10^{-3}$, approximates a coin flip in predicting membership.
Our results, detailed in~\autoref{tab:vulnerable_num}, indicate that 627 samples (approximately 2.5\%) in CIFAR-10 can be chosen as vulnerable samples. We noticed most of the remaining samples (16,586 out of 25,000 samples) are less memorized and difficult for an attacker to infer membership. We consider data samples other than these 627 samples as safe and protected ones. As expected, 2,583 samples (approximately 10.33\%) in CIFAR-100 are highly memorized. Conversely, we identified 16,586 less memorized samples in CIFAR-10 and 6,639 in CIFAR-100.

\begin{table}[!h]
\footnotesize
\caption{MIA's TPR@FPR and number of vulnerable samples (256 shadow models on one-round membership inference).}\vspace{-0.1in}
\begin{tabular}{l|cccc}
\hline
Dataset & \begin{tabular}[c]{@{}c@{}}TPR@\\ 0.1\%FPR\end{tabular} & \begin{tabular}[c]{@{}c@{}}TPR@\\ 0.01\%FPR\end{tabular} & \begin{tabular}[c]{@{}c@{}}\#Highly\\ memorized\end{tabular} & \begin{tabular}[c]{@{}c@{}}\#Less\\ memorized\end{tabular} \\ \hline
CIFAR-10 & 4.1\% & 2.5\% & 627 & 16,586\\
CIFAR-100 & 24.33\% & 10.33\% & 2,583 & 6,639\\ \hline
\end{tabular}
\vspace{-0.15in}
\label{tab:vulnerable_num}
\end{table}

\subsubsection{Target Data Indistinguishability}
\noindent\textbf{Image.} To ensure the reliability of membership inference evaluations, it is critical that the selected two cases of target data do not exhibit distribution shifts and are resistant to \textit{blind attacks}~\cite{das2024blind}. In blind attacks, the adversary does not rely on model training and instead attempts to distinguish data solely based on its distributional characteristics. We assess the indistinguishability of members (\textit{Unlearned}) and non-members (\textit{Out/Held-out}) samples using two metrics: Label Overlap, which quantifies the proportion of shared classes to rule out label-based separability; and Embedding Classifier, which trains a linear classifier from embeddings extracted from a ViT-based model to distinguish the two sets. A near-chance classifier accuracy ($\approx$50\%) indicates no clear distributional difference. We also include a uniformly random selection baseline to reflect the expected distinguishability when selecting 250 samples from each set—particularly. Results in \autoref{tab:blind-attack} show both canary (our targets) and random selections yield similar overlap and indistinguishability.

\begin{table}[h]
\centering
\footnotesize
\caption{Blind attack results: label overlap and classifier accuracy.}
\label{tab:blind-attack}
\vspace{-0.1in}
\begin{tabular}{@{}lcccc@{}}
\toprule
\multirow{2}{*}{} & \multicolumn{2}{c}{Canary (ours)} & \multicolumn{2}{c}{Random} \\ \cmidrule(lr){2-3} \cmidrule(lr){4-5}
 & \begin{tabular}[c]{@{}c@{}}Labels \\ Overlap\end{tabular} & \begin{tabular}[c]{@{}c@{}}Accuracy \\ (Classifier)\end{tabular} & \begin{tabular}[c]{@{}c@{}}Labels \\ Overlap\end{tabular} & \begin{tabular}[c]{@{}c@{}}Accuracy \\ (Classifier)\end{tabular} \\ \midrule
CIFAR-10 & 92.0\% & 52.0\% & 91.6\% & 51.3\% \\
CIFAR-100 & 87.2\% & 51.3\% & 87.6\% & 50.0\% \\
TinyImageNet & 54.8\% & 53.3\% & 53.2\% & 46.7\% \\ \bottomrule
\end{tabular}
\end{table}

\noindent \textbf{Text}. We randomly select the \textit{out} and \textit{unlearned} target sequences from the test set of WikiText-103, ensuring that both groups are drawn from the same underlying distribution. Since our membership inference only queries the final 7 tokens of each sample, we evaluate whether these target spans are lexically indistinguishable from one another. We compute both exact 7-gram matches and token-wise (1-gram) overlap between the out and unlearned spans under different selection modes: first 7 tokens, final 7 tokens, and random 7 tokens. Exact matches are rare in all settings ($\leq$ 2\%), but we observe relatively high token-wise overlap for the final 7 tokens (mean: 54.3\%), which we use as our targets.

\section{Empirical evidence of mismatch between  \textit{Out} and \textit{Held-out} distributions}
\begin{figure}[!ht]
\footnotesize
    \centering
    \begin{subfigure}[b]{\linewidth}
        \centering
        \includegraphics[width=\linewidth]{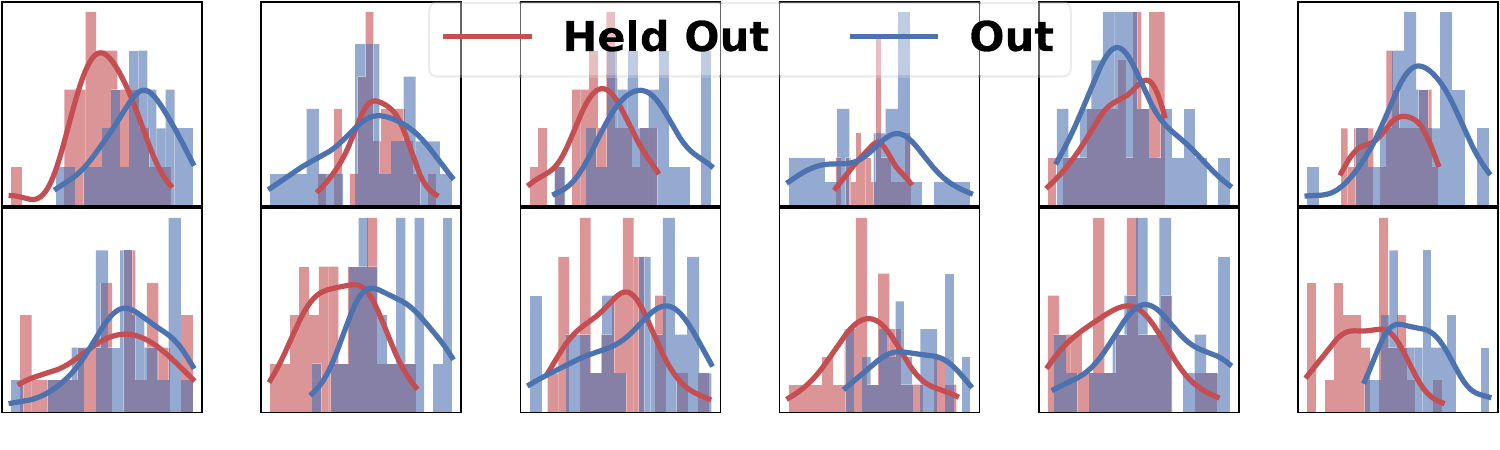}
        \caption{$\ell_1$ Sparse}
        \label{fig:random_Q_scrub}
    \end{subfigure}
    \begin{subfigure}[b]{\linewidth}
        \centering
        \includegraphics[width=\linewidth]{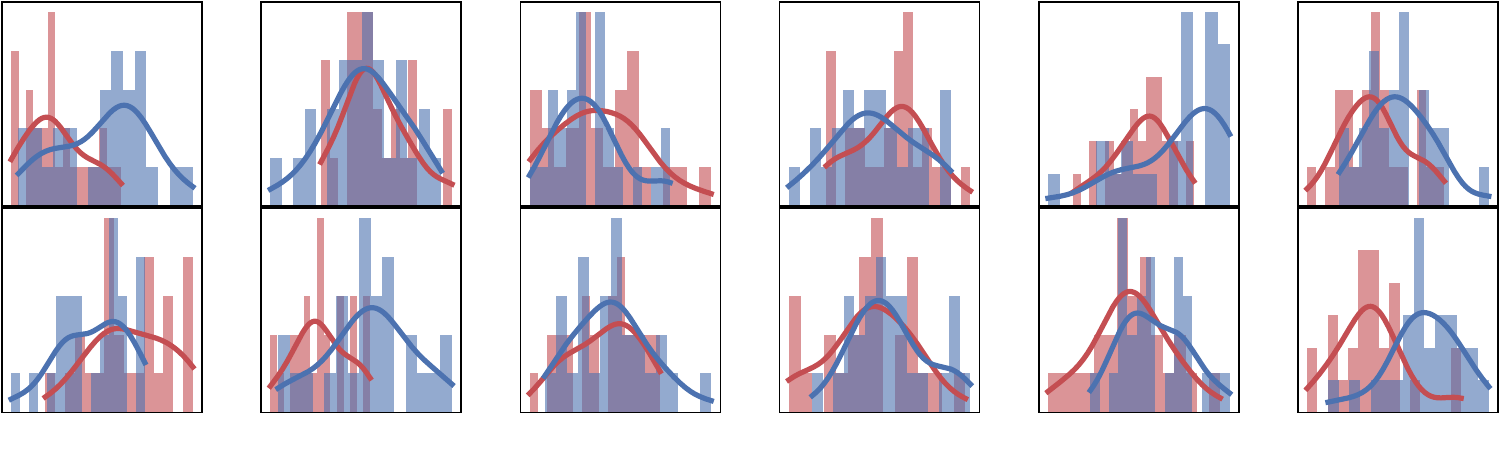} 
        \caption{Scrub}
        \label{fig:random_Q_FT}
    \end{subfigure}
    \begin{subfigure}[b]{\linewidth}
        \centering
        \includegraphics[width=\linewidth]{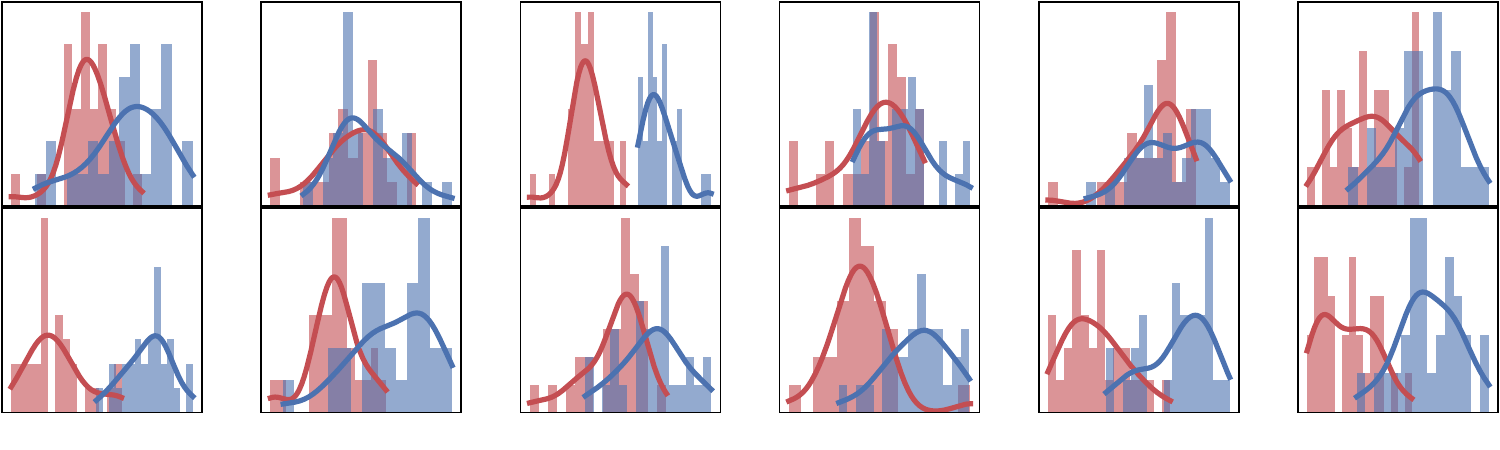} 
        \caption{GA+}
        \label{fig:random_Q_GA}
    \end{subfigure}
    \begin{subfigure}[b]{\linewidth}
        \centering
        \includegraphics[width=\linewidth]{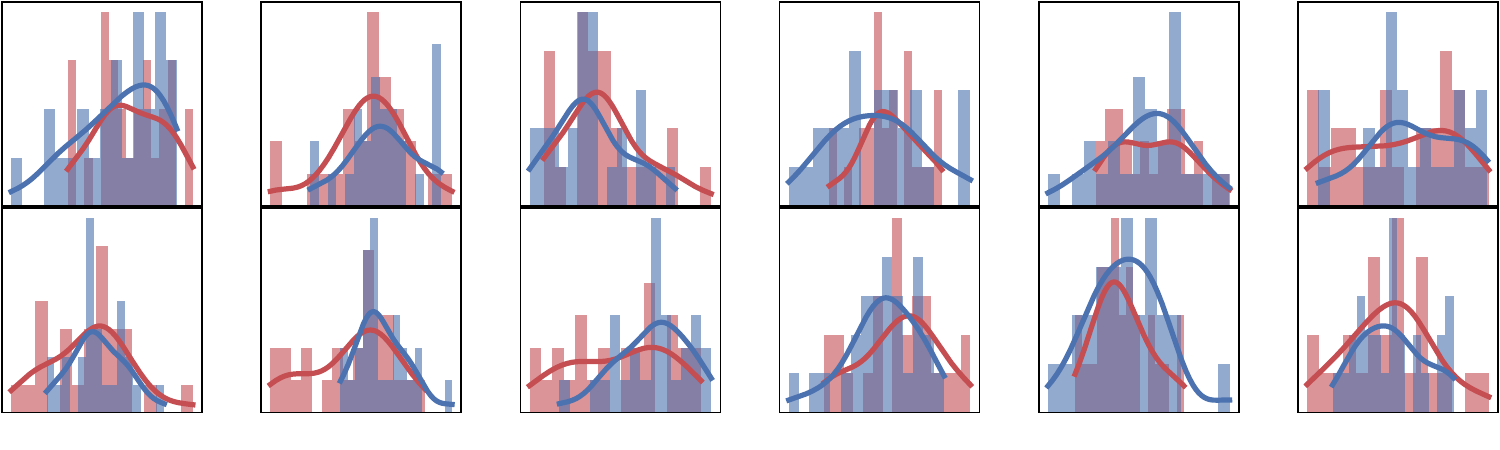} 
        \caption{Retrain}
        \label{fig:random_Q_retrain}
    \end{subfigure}
\caption{Logit-scaled confidences of two distributions held-out ($\mathcal{Q}_h$) and out ($\mathcal{Q}_R$) from 16 randomly samples of class-5 of CIFAR-10 (compatible with the U-LiRA setting). Our empirical evidences are showing that $\mathcal{Q}_h$ is not similar to $\mathcal{Q}_u$ unless in Retraining.}
\label{Q_U_LIRA}
\end{figure}

To demonstrate the distinguishability of $\mathcal{Q}_h$ and $\mathcal{Q}_R$, and to support our claim in Section~\ref{ulira-limit} that these two distributions are not similar for all samples, we present logit-scaled confidences derived from our shadow models representing these distributions. We selected 12 random samples from class 5 of the CIFAR-10 dataset and utilized 90 shadow models for this analysis. The logit-scaled confidences were evaluated using $\ell_1$ Sparse, GA+, NegGrad+, and Retrain (standard) benchmarks, as shown in~\autoref{Q_U_LIRA}. In the figure, the red distribution represents cases where the sample was excluded from both shadow model training and shadow model unlearning (held-out), while the blue distribution represents cases where the sample was excluded from training alone (out). Our results demonstrate that for inexact unlearning benchmarks, such as $\ell_1$ Sparse, GA+, and NegGrad+, the held-out and out distributions differ significantly across all 12 samples. In contrast, when employing the Retrain (standard) benchmark, these distributions appear similar, suggesting consistency. This discrepancy highlights a key limitation of inexact unlearning methods, where $\mathcal{Q}_h \not\sim \mathcal{Q}_R$ directly challenges the assumption that these distributions are interchangeable. This finding emphasizes the need for more nuanced evaluation frameworks to account for such distributional differences in unlearning.

\vspace{-0.3in}

\section{Discussions}

\subsection{Possible Solutions to Mitigate Privacy Risk of Inexact Unlearning Designs} 

We discuss two possible solutions to mitigate privacy leakage of inexact unlearning.

\vspace{0.05in}

\noindent\textbf{Estimating Memorization}. To mitigate the negative effects of batch optimizations during unlearning, one approach is to estimate the memorization levels of all forget samples and sequentially unlearn similar samples within the same round. However, accurately assessing per-sample memorization is computationally expensive, often requiring the training of numerous models. Prior work has shown that many of the most vulnerable samples tend to be mislabeled or ambiguous, suggesting that model trainers could use static analysis techniques to pre-identify and handle these cases separately. As a proof of concept, we assume that the model provider has some means of estimating per-sample memorization and can construct mini-batches ranked by the memorization scores of the forget data. Using this strategy, we observe a notable reduction in attack performance: attack accuracy drops to 61\%, and TPR@FPR 1\% decreases to 3\% on the Scrub unlearning.

\vspace{0.05in}

\noindent\textbf{Training with DP-SGD.} Differentially Private SGD (DP-SGD)~\cite{dpsgd} is a principled training framework that provides worst-case privacy guarantees for every individual sample. This enables provable, per-sample privacy estimates~\cite{gradients_looklike}, which can be leveraged to assess the privacy status of forget samples. Such a property makes DP-trained models a compelling candidate for unlearning, as one can, in theory, show that the effective privacy loss ($\epsilon$) for a forgotten sample is near zero. However, DP training is not commonly adopted in practice due to its privacy-utility trade-offs \cite{FengMWLQH24}. Despite this, designing unlearning algorithms specifically for DP-trained models remains a promising direction. Since DP-trained models are often meant to protect sensitive data, effective unlearning in this setting could be both feasible and beneficial. Exploring unlearning techniques that push the per-sample $\epsilon$ lower (ideally towards zero) could offer a new perspective on practical, certifiable unlearning.

\subsection{Notes on Provided Results}

\noindent\textbf{Sensitivity to Hyperparameters.}
We observe that unlearning algorithms are significantly more sensitive to hyperparameter settings than standard training, especially when forgetting high-risk canaries. While we perform grid search to select reasonable settings, small variations can lead to meaningful shifts in attack performance. For example, applying the CIFAR-10 unlearning setting to CIFAR-100—by reducing just one unlearning step—raises the TPR@1\%FPR to 24.5\%. This suggests that even subtle changes can disturb unlearning performance and even possibly amplify privacy leakage.

\vspace{0.05in}

\noindent\textbf{Varying Canary Rates.}
One can vary the canary rates without retraining the shadow models. In our default setting (Section~\ref{canary_setting}), nearly 55\% of the forget set consists of vulnerable samples canaries. As an illustrative example, we reduce this rate to 5\% by injecting more non-canary samples into forget data, and adjust the Scrub unlearning hyperparameters accordingly. The attack remains effective. On CIFAR-10, we observe an average TPR@1\%FPR of 8.7\%, TPR@5\%FPR of 26.6\%, and attack accuracy of 68.5\%. On TinyImageNet, TPR@1\%FPR reaches 9.6\%, TPR@5\%FPR is 28.2\%, and accuracy is 65.2\%. These results demonstrate that \sys's effectiveness at a lower canary rate.

%% file: Tables/baseline_param.tex

\begin{table}[h]
\centering
\caption{Hyper-parameters in unlearning benchmarks.}
\vspace{-0.1in}
\scriptsize
\resizebox{\columnwidth}{!}{%
\begin{tabular}{@{}l|l@{}}
\toprule
\begin{tabular}[c]{@{}l@{}} \end{tabular} & Unlearning Tunable Hyper-parameters \\ \midrule
\multicolumn{1}{l|}{$\ell_1$-sparse} & \begin{tabular}[c]{@{}l@{}}learning rate: \{0.01\}, sparsity parameter ($\alpha$): \{$10^{-4}$, $2 \times 10^{-4}$, $5 \times 10^{-4}$, $3 \times 10^{-4}$ \}, \\ batch size: \{16, 64, 128\}, sparsity scheduler: \{linear increase, decay, constant\}\end{tabular} \\ \midrule
\multicolumn{1}{l|}{Scrub} & \begin{tabular}[c]{@{}l@{}}learning rate: \{$5 \times 10^{-4}, 10^{-4}, 5 \times 10^{-5}\}$, $\alpha$: 0.1, $\beta$: 0.0, $\gamma$: 0.99,\\ forget batch size: \{16, 64, 128\}, retain batch size: \{16, 64, 128\}, \\ maximizing steps: \{1, 2, 4, 6\}, minimizing steps: \{2, 5, 6, 10\}\end{tabular} \\ \midrule
\multicolumn{1}{l|}{GA+} & \begin{tabular}[c]{@{}l@{}}learning rate: \{$10^{-4}$, $10^{-3}$\}, unlearn epochs: \{3, 5, 7, 10, 15, 20\}, \\ refine epochs: \{0, 3, 4, 5\}, batch size: \{16, 64, 128\}\end{tabular} \\ \midrule
\multicolumn{1}{l|}{NegGrad+} & \begin{tabular}[c]{@{}l@{}}learning rate: $5 \times 10^{-4}$, $5 \times 10^{-5}$, $\alpha$: \{0.1, 0.3, 0.35, 0.4, 0.5, 0.9\}, \\ forget batch size: \{16, 64, 128\}, retain batch size: \{16, 64, 128\}, \\ maximizing steps: \{1, 2, 4, 6\}\end{tabular} \\ \bottomrule
\end{tabular}
\label{tab:unleanring-baselines-param}
}\vspace{-0.1in}
\end{table}

\normalsize